\documentclass[12pt,preprint]{aastex}
\usepackage{natbib}

\def\chandra{{\it Chandra\/}}
\def\rosat{{\it ROSAT\/}}

\begin{document}
\title{Deep-Survey Constraints on \hbox{X-ray} Outbursts from Galactic Nuclei}

\author{B.~Luo,\altaffilmark{1}
W.~N.~Brandt,\altaffilmark{1}
A.~T.~Steffen,\altaffilmark{1}
\& F.~E.~Bauer\altaffilmark{2}
}

\altaffiltext{1}{Department of Astronomy \& Astrophysics, 525
Davey Lab, The Pennsylvania State University, University Park, PA
16802, USA} \altaffiltext{2}{Columbia Astrophysics Laboratory,
Columbia University, Pupin Labortories, 550 W. 120th St., Rm 1418,
New York, NY 10027, USA}

\begin{abstract}
Luminous \hbox{X-ray} outbursts with variability amplitudes as
high as $\sim 1\,000$ have been detected from a small number of
galactic nuclei. These events are likely associated with transient
fueling of nuclear supermassive black holes. In this paper, we
constrain \hbox{X-ray} outbursts with harder spectra, higher
redshifts, and lower luminosities than have been studied
previously. We performed a systematic survey of $24\,668$ optical
galaxies in the \chandra\ Deep Fields to search for such
\hbox{X-ray} outbursts; the median redshift of these galaxies is
$\sim 0.8$. The survey spans $798$ days for the \chandra\ Deep
Field-North, and $1\,828$ days for the \chandra\ Deep Field-South.
No outbursts were found, and thus we set upper limits on the rate
of such events in the Universe, which depend upon the adopted
outburst \hbox{X-ray} luminosity. For an outburst with
\hbox{X-ray} luminosity $\ga 10^{43}$~ergs~${\rm s}^{-1}$ and a
duration of 6 months, the upper limit on its event rate is $\sim
10^{-4}$~galaxy$^{-1}$~yr$^{-1}$, roughly consistent with
theoretical predictions. Compared to previous survey results, our
harder-band and deeper survey suggests that the outburst rate may
increase by a maximum factor of 10 when considering both obscured
\hbox{X-ray} outbursts and redshift evolution from $z\sim 0$ to
$z\sim 0.8$. Our results also suggest that the \hbox{X-ray}
luminosity function for moderate-luminosity active galactic nuclei
is not primarily due to stellar tidal disruptions.
\end{abstract}
\keywords{galaxies: active --- galaxies: nuclei --- X-rays}

\section{INTRODUCTION}

X-ray observations, mainly with \rosat, have found about six
transient, large-amplitude \hbox{X-ray} outbursts from galactic
nuclei (e.g., Donley et~al. 2002; Komossa 2002; Komossa et~al.
2004; Vaughan, Edelson, \& Warwick 2004; and references therein).
These events have variability factors of \hbox{$\approx
20$--1\,000}, peak $0.5-2$~keV \hbox{X-ray} luminosities
comparable to those of local Seyfert galaxies \hbox{($\approx
10^{42}$--$10^{44}$~erg~s$^{-1}$)}, decay timescales of months,
and soft \hbox{X-ray} spectra. They have been observed in both
active and inactive galaxies (i.e., galaxies with and without a
persistently accreting supermassive black hole, respectively). The
estimated event rates of these outbursts suffer from large
systematic and statistical uncertainties. In nearby inactive
galaxies the event rate is $\sim 10^{-5}$~galaxy$^{-1}$~yr$^{-1}$,
while in active galaxies the event rate appears to be $\sim 100$
times higher (e.g., Donley et~al. 2002). At least some of these
outbursts induce accompanying optical nuclear variability (e.g.,
Brandt, Pounds, \& Fink 1995; Grupe et~al. 1995). Recently, five
new candidate \hbox{X-ray} outbursts were found in the {\it
XMM-Newton} Slew Survey \citep{Esquej2007}, and one outburst has
been detected in the ultraviolet and is thought to have the same
origin as \hbox{X-ray} outbursts \citep{Gezari2006}.

The physical origin of nuclear \hbox{X-ray} outbursts remains
mysterious. The most likely explanation is that they are caused by
transient fueling events of nuclear supermassive black holes
(SMBHs). Since SMBHs are thought to be ubiquitous in nucleated
galaxies (e.g., Ferrarese \& Ford 2005 and references therein),
fueling events seem inevitable in crowded galactic centers.
Fueling may occur when a star, planet, or gas cloud is tidally
disrupted and partially accreted (e.g., Rees 1990; Ulmer 1999; Li,
Narayan, \& Menou 2002). The predicted event rate for stellar
tidal disruptions \citep[e.g.,][]{Magorrian1999,Syer1999} is
roughly consistent with the poorly constrained rate of
large-amplitude \hbox{X-ray} outbursts in inactive galaxies.
However, \citet{Wang2004} predicted a rate that was a factor of
\hbox{$\sim 10$} higher than earlier results, mainly due to
downwardly revised black-hole masses. This predicted rate,
apparently in excess of the observed rate of \hbox{X-ray}
outbursts, may indicate that a large fraction of tidal-disruption
events do not exhibit expected \hbox{X-ray} outburst
characteristics, perhaps due to short durations of the events or
\hbox{X-ray} obscuration. In some cases, transient fueling could
be due to accretion-disk instabilities (e.g., Siemiginowska,
Czerny, \& Kostyunin 1996). Some outbursts could also perhaps be
explained by \hbox{X-ray} afterglows of gamma-ray bursts (GRBs),
though no simultaneous GRBs were reported for the known
\hbox{X-ray} outbursts \citep[e.g.,][]{Komossa99}.

Aside from the innate scientific interest of nuclear \hbox{X-ray}
outbursts, determining the rate and properties of such events is
relevant to planning future missions such as the {\it Black Hole
Finder Probe (BHFP)\/}, {\it Lobster\/}, and {\it eROSITA\/}. The
{\it BHFP\/}, for example, is likely to observe in the hard
\hbox{X-ray} band, and the large sky coverage of this mission will
allow intrinsically rare transient events to be studied (e.g.,
Grindlay 2005). Facilities such as the {\it Large Synoptic Survey
Telescope\/}, the {\it Joint Dark Energy Mission\/}, and the {\it
Laser Interferometer Space Antenna\/} should also allow the
accompanying optical (e.g., Brandt 2005) and gravitational-wave
(e.g., Kobayashi et~al. 2004) outbursts to be studied.

Previous studies of \hbox{X-ray} outbursts from galactic nuclei
have delivered fascinating results but require extension so that
this phenomenon can be better understood. For example, advances
are required in the following directions:

\begin{enumerate}

\item Outbursts with harder \hbox{X-ray} spectra. As noted above,
the outbursts discovered to date have generally had soft
\hbox{X-ray} spectra with effective \hbox{0.2--2.0~keV} power-law
photon indices of \hbox{$\Gamma\approx 3$--5}. However, this may
partially be a selection effect owing to the soft \hbox{X-ray}
bandpass of the \rosat\ satellite. Transient fueling events of
SMBHs should be capable of generating harder \hbox{X-ray} spectra
\hbox{($\Gamma\approx 1.7$--2.2)} via Compton upscattering in the
accretion flow; such spectra are observed from most active
galactic nuclei (AGNs) as well as from transient Galactic black
holes. Indeed, one \hbox{X-ray} outburst has shown evidence for
spectral hardening as it declined (Komossa et~al. 2004). In
addition, obscuration can harden the observed \hbox{X-ray} spectra
of an outburst, which would make it more difficult to detect in
the current outburst surveys. A fairly small column density of
$N_{\rm H}=5\times 10^{21}$~cm$^{-2}$ reduces the expected
\hbox{0.2--2.0~keV} flux from a soft-spectrum outburst by a factor
of $\sim 20$.

\item Higher redshift outbursts. The \hbox{X-ray} outbursts
discovered to date are at low redshift with \hbox{$z=0.01$--0.15}.
Since only a small fraction of cosmic time is spanned by this
redshift range and the source statistics are limited, little is
known about the redshift evolution of their frequency. It is
plausible that the \hbox{X-ray} outburst rate could show evolution
over cosmic time, considering that AGNs and galaxies both show
strong evolution. The number densities of comparably \hbox{X-ray}
luminous AGNs evolve upward by factors of $\approx 10$--30 out to
$z\sim 1$ (e.g.,  Brandt \& Hasinger 2005 and references therein),
and Milosavljevi\'c, Merritt, \& Ho (2006) have suggested that a
significant portion of the \hbox{X-ray} luminosity function of
such AGNs may be comprised of sources powered by tidal
disruptions.

\item Lower luminosity outbursts. Transient phenomena in complex
natural systems often follow power-law distributions in frequency
of occurrence, such that low-amplitude events are more common than
high-amplitude events (e.g., earthquakes, avalanches, solar
flares, and ecological extinctions). It is plausible that such a
distribution applies for \hbox{X-ray} outbursts in galactic
nuclei. Low-mass fueling events, such as partial tidal disruptions
or the accretion of brown dwarfs, planets, or small gas clouds,
would then be more common than high-mass fueling events.

\end{enumerate}

\noindent In this paper, we utilize data from the \chandra\ Deep
Fields (see Brandt \& Hasinger 2005 for a review) to constrain the
rate of \hbox{X-ray} outbursts with harder \hbox{X-ray} spectra,
higher redshifts, and lower luminosities than those studied to
date. Observations of the \chandra\ Deep Fields were made in
several well-separated and individually sensitive ``epochs,''
allowing effective constraints to be placed upon \hbox{X-ray}
outbursts that evolve on timescales of months in the rest frame.
Some variability work has been performed on the \chandra\ Deep
Field sources. For example, Paolillo et~al. (2004) studied the
\hbox{X-ray} variability of $\sim 350$ sources detected in the
\chandra\ Deep Field-South (\hbox{CDF-S}), but this work has not
been optimized to detect and to constrain systematically
\hbox{X-ray} outbursts of the type relevant here.

We adopt $H_0=70$~\hbox{km s$^{-1}$ Mpc$^{-1}$}, $\Omega_{\rm M}$
= 0.3, and $\Omega_{\Lambda}$ = 0.7 throughout this paper. All
coordinates are J2000. The relevant Galactic column densities are
$1.3\times 10^{20}$~cm$^{-2}$ for the \chandra\ Deep Field-North
(\hbox{CDF-N}; Lockman 2004) and $8.8\times 10^{19}$~cm$^{-2}$ for
the \hbox{CDF-S} (Stark et~al. 1992).
All \hbox{X-ray} fluxes and luminosities quoted throughout this
paper have been corrected for Galactic absorption using these
column densities.

\section{THE SURVEY}

In this survey, we focused on optically detected galaxies in the
\hbox{CDF-N} \citep{Brandt2001,Alexander2003} and the \hbox{CDF-S}
\citep{Giacconi2002,Alexander2003}. An optical galaxy sample was
compiled from several catalogs, with the redshifts collected from
literature. We grouped the \chandra\ observations into several
well-separated ``epochs''. The five epochs for the \hbox{CDF-N}
span \hbox{$798$} days, and the four epochs for the \hbox{CDF-S}
span \hbox{$1\,828$} days. At the median redshift of \hbox{$\sim
0.8$} for the galaxies in our sample, the rest-frame coverage
would be \hbox{$443$} days for the \hbox{CDF-N} and
\hbox{$1\,016$} days for the \hbox{CDF-S}. \hbox{X-ray} count
rates or upper limits for every optical galaxy were measured for
each epoch in three standard \chandra\ bands: \hbox{0.5--8.0}~keV
(full band; FB), 0.5--2.0~keV (soft band; SB), and 2--8~keV (hard
band; HB). These measurements were then analyzed to search for any
\hbox{X-ray} outburst candidates. Here we define ``\hbox{X-ray}
outbursts'' to be transient events in galaxy nuclei that cause the
count rate to vary by a minimum factor of 20 in one of the three
standard bands. A minimum variability factor of 20 was chosen to
discriminate against normal AGN variability, which typically has
variability factors of \hbox{$\approx$2--5} but can be as large as
\hbox{$\approx$10--15} (e.g., Paolillo et~al. 2004).

\subsection{Optical Galaxy Sample}
Optical galaxies were selected from the catalog of
\citet{Capak2004} for the \hbox{CDF-N}, and from the COMBO-17
catalog \citep{Wolf2004} for the \hbox{CDF-S}. To ensure that
these galaxies were covered by either the \hbox{CDF-N} or the
\hbox{CDF-S}, we used the 0.5--8.0~keV exposure maps of the 2 Ms
\hbox{CDF-N} and 1 Ms \hbox{CDF-S} as filters. A small portion
($\sim30~{\rm arcmin}^2$) of the \hbox{CDF-N} was not covered by
the Capak catalog, so the coverage of this survey is $\sim$418
${\rm arcmin^2}$ for the \hbox{CDF-N} and $\sim$391 ${\rm
arcmin^2}$ for the \hbox{CDF-S}. We only chose optical galaxies
with AB magnitudes $R<25$, to optimize our selection of a large
galaxy sample with redshift measurements. We examined manually
those sources with $R<22$ and removed some false sources, which
were generally around bright objects. Galaxies in the COMBO-17
catalog with values for photometry flags $\ge8$ were ignored, as
suggested by the creators of that catalog. In this way, we
selected $13\,699$ galaxies in the \hbox{CDF-N} and $11\,077$
galaxies in the \hbox{CDF-S}.

\begin{figure}
\centerline{\includegraphics[scale=0.5]{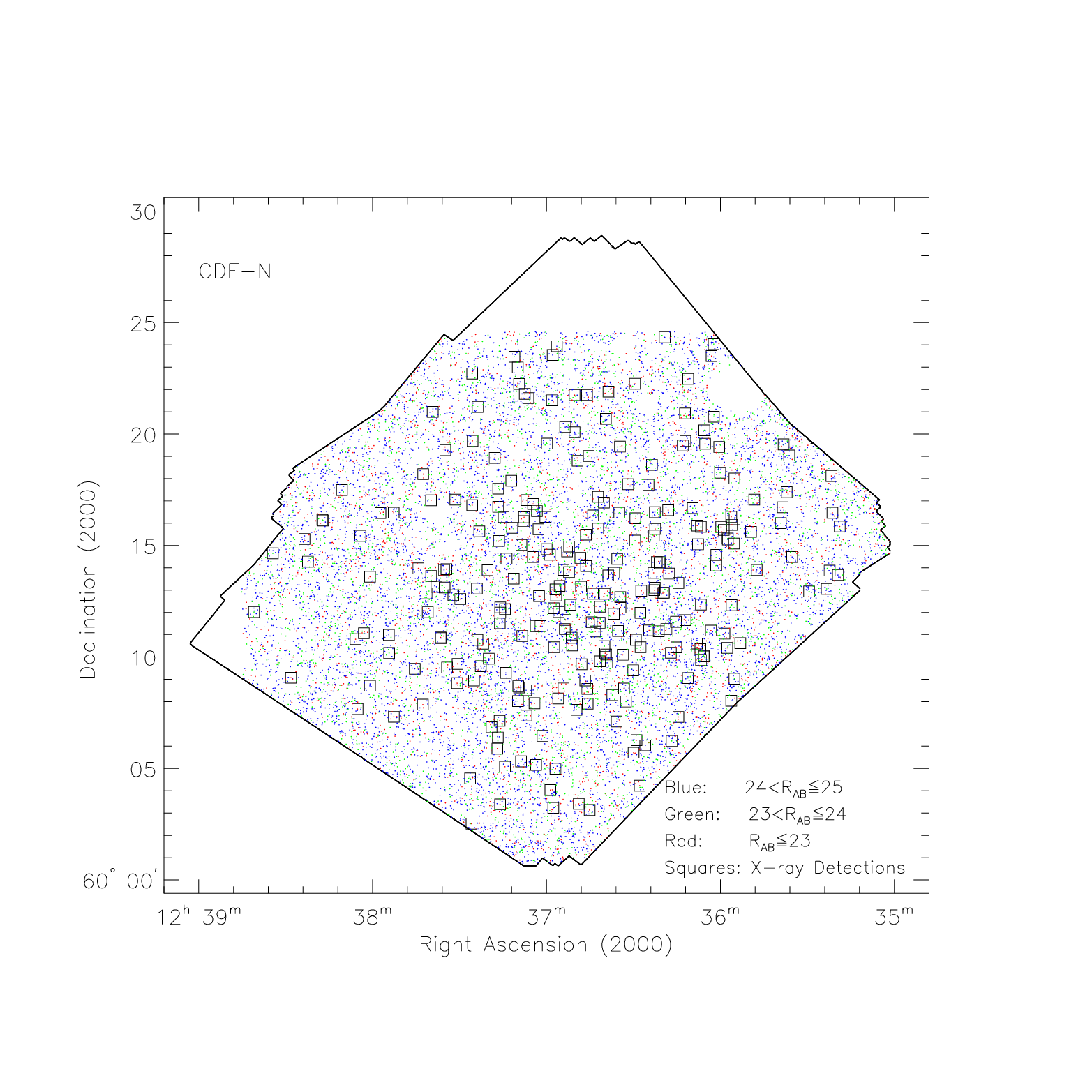}
\includegraphics[scale=0.5]{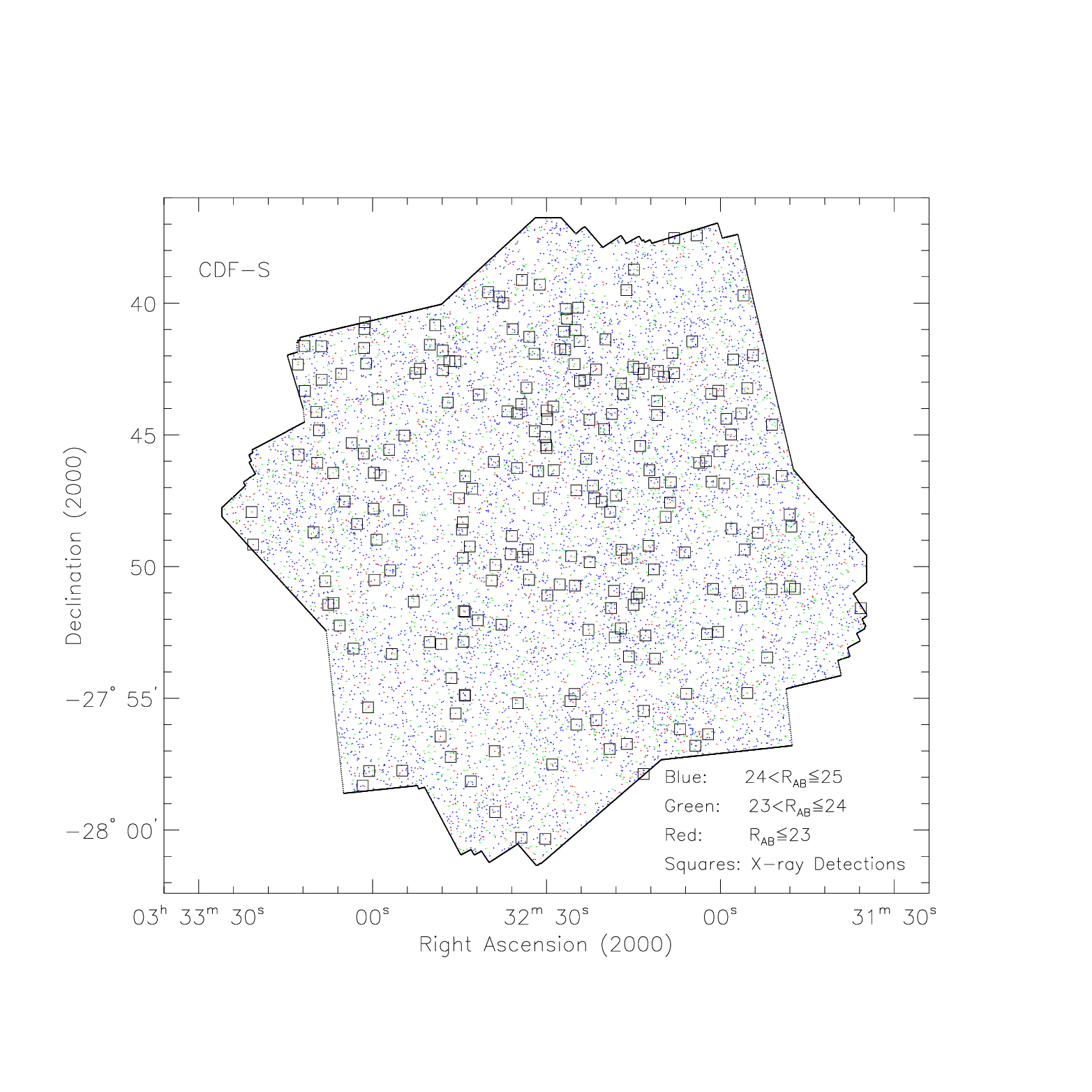}}
\figcaption{Spatial distributions of the galaxies in this survey
in the \hbox{CDF-N} and \hbox{CDF-S}. Dots with different colors
represent galaxies in different magnitude bins. Squares are
galaxies with \hbox{X-ray} counterparts as listed in Table
\ref{tblepoch}. The gap at the top of the \hbox{CDF-N} is due to
the lack of coverage of the optical catalog. Color figures are
available online. \label{region}}
\end{figure}

\begin{figure}
\centerline{\includegraphics[scale=0.68]{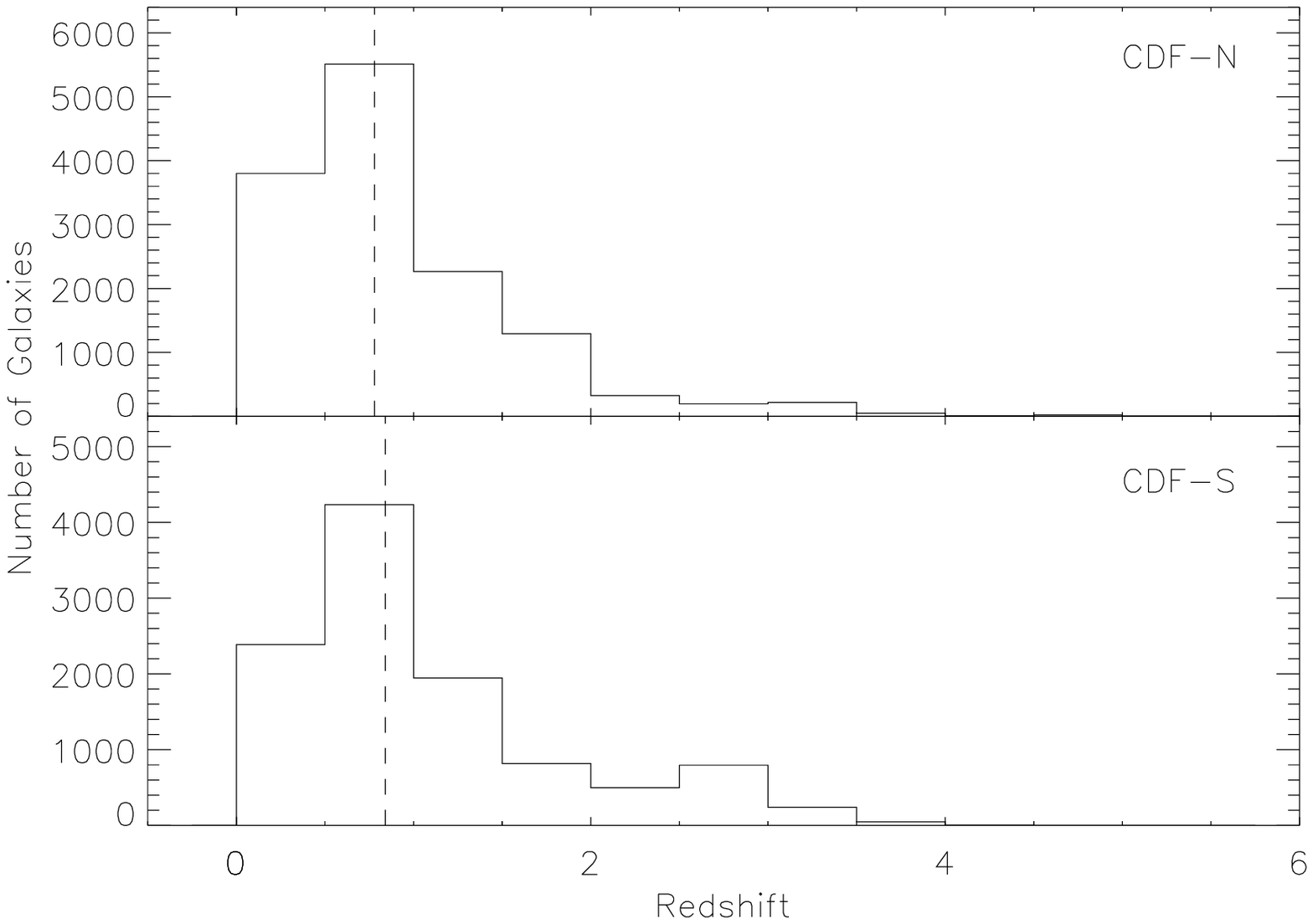}}
\figcaption{Redshift distributions of the galaxies in this survey
for the \hbox{CDF-N} and \hbox{CDF-S}. The median and mean
redshifts are 0.78 and 0.91 for the \hbox{CDF-N}, and 0.84 and
1.06 for the \hbox{CDF-S}, respectively. The dashed lines show the
median values. \label{zdis}}
\end{figure}

We searched the literature for redshift measurements for this
optical galaxy sample. Spectroscopic redshifts were obtained from
the spectroscopic surveys of \citet{Cowie2004}, \citet{Reddy2006},
and the ``Team Keck'' Treasury Redshift Survey \citep{Wirth2004}
in the \hbox{CDF-N}, and from the VIMOS VLT Deep Survey
\citep{LeFevre2004} and the VLT/FORS2 spectroscopic survey
\citep{Vanzella2005,Vanzella2006} in the CDF-S. Photometric
redshifts were obtained from \citet{Capak2007inprep} in the
\hbox{CDF-N}, and from \citet{Mobasher2004} and \citet{Wolf2004}
in the \hbox{CDF-S}. A matching radius of $1^{\prime\prime}$ was
used when cross correlating different optical catalogs. We did not
find redshift measurements for 108 faint galaxies in the
\hbox{CDF-S}. After removing these sources our final optical
galaxy sample contained $13\,699$ and $10\,969$ galaxies in the
\hbox{CDF-N} and \hbox{CDF-S}, respectively. The spatial
distributions of these galaxies in the \hbox{CDF-N} and
\hbox{CDF-S} are shown in \hbox{Figure \ref{region}}. We found
spectroscopic redshifts for $2\,100$ galaxies in the \hbox{CDF-N}
and $1\,534$ galaxies in the \hbox{CDF-S}, with the remaining
sources having photometric redshifts. The redshift distributions
are shown in \hbox{Figure \ref{zdis}}. To test the reliability of
the photometric redshifts, we examined sources with both
photometric and spectroscopic redshift measurements, and we
studied the offset distribution of these two types of redshifts in
the \hbox{CDF-N} and \hbox{CDF-S} separately. In both fields, the
median offsets approach 0. The interquartile ranges are only 0.11
in the \hbox{CDF-N} and 0.20 in the \hbox{CDF-S}. The ``stellarity
index'' ({\sc class\_star} parameter in SExtractor) is a robust
galaxy identifier, with 0 for confirmed galaxies, and 1 for
confirmed stars \citep[e.g.,][]{Groenewegen2002}. Less than 2.5\%
of these galaxies had stellarity indices $\ge 0.9$, so stars were
well removed from this sample.

This sample of $24\,668$ galaxies contains a representative mix of
field galaxies. Only $\sim 2\%$ of the galaxies are \hbox{X-ray}
detected, and thus the AGN fraction is small.
\citet{Lehmer2007inprep} constructed a similar galaxy sample in
the \chandra\ Deep Fields, with $z_{850}<23$ and
\hbox{$z=0$--1.4}. They studied the rest-frame optical colors as
well as the S\'{e}rsic indices \citep[e.g.,][]{Haussler2007} of
galaxies in their sample, finding $2\,544$ late-type galaxies and
$727$ early-type galaxies out of $3\,271$ galaxies. Our sample
should have a similar composition in terms of galaxy morphology.
Recent close-pair studies have suggested that the merger rate of
field galaxies remains fairly small out to $z\sim 1$
\citep[e.g.,][$\sim 7\%$ infrared-selected pairs]{Bundy2004}.
\citet{Elmegreen2007} examined the GEMS (Galaxy Evolution from
Morphology and SEDs) and the southern GOODS (Great Observatories
Origins Deep Survey) fields, which overlap with the \hbox{CDF-S},
and identified $\sim 300$ interacting galaxies to $z\sim 1.4$ out
of over $8\,000$ optical galaxies. Thus our sample should also be
dominated by non-interacting galaxies.

\subsection{\hbox{X-ray} Data}

X-ray data analyses were performed on the \hbox{CDF-N},
\hbox{CDF-S}, and the Extended \chandra\ Deep Field-South
\citep[\hbox{E-CDF-S};][]{Lehmer2005}. There are 20 observations
for the \hbox{CDF-N}, 11 observations for the \hbox{CDF-S}
\citep{Alexander2003}, and 9 observations for the \hbox{E-CDF-S}
\citep{Lehmer2005}. We grouped these into ``epochs'' according to
their observation dates, five epochs for the \hbox{CDF-N} and four
epochs for the \hbox{CDF-S} (see Table \ref{tblepoch}). The
\hbox{E-CDF-S} observations took place $\sim4$ years after the
last \hbox{CDF-S} observation and were counted as the fourth epoch
for the \hbox{CDF-S}. The \hbox{E-CDF-S} covered the entire
\hbox{CDF-S} and consisted of four distinct observational fields
with different observation dates. We list these fields separately
in Table~\ref{tblepoch}. We created images of each epoch from the
reduced and cleaned level 2 event files
\citep{Alexander2003,Lehmer2005}, using the standard {\it ASCA}
grade set ({\it ASCA} grades 0, 2, 3, 4, and 6) for the three
standard bands. There were thus 15 \hbox{CDF-N} images, 9
\hbox{CDF-S} images, and 12 \hbox{E-CDF-S} images. The aim point
for a given image was taken to be the average value over its
observations weighted by exposure time.

To construct \hbox{X-ray} source lists, we ran {\sc wavdetect}
\citep{Freeman2002} on the 24 images in the \hbox{CDF-N} and
\hbox{CDF-S} using a ``$\sqrt{2}$ sequence'' of wavelet scales
(i.e., 1, $\sqrt{2}$, 2, 2$\sqrt{2}$, 4, 4$\sqrt{2}$, and 8
pixels). The false-positive probability threshold in each {\sc
wavdetect} run was set to $1\times 10^{-6}$. Source lists for the
\hbox{E-CDF-S} were taken directly from \citet{Lehmer2005} because
their source-detection method is the same as that used here. To
improve the accuracy of the \hbox{X-ray} source positions, we
matched the optical and \hbox{X-ray} source lists with a matching
radius of $2\farcs 5$ and centered the distributions of offsets in
right ascension and declination between the optical and
\hbox{X-ray} source positions. This resulted in small ($<1\farcs
0$) image-dependent astrometric shifts for all \hbox{X-ray}
sources. Compared to the main {\it Chandra} \hbox{X-ray} source
catalogs in \citet{Alexander2003}, our \hbox{X-ray} source lists
give $\sim 10$ new sources for each image, due to the
less-conservative false-positive probability threshold we were
using or the variability of some sources.

We are interested in the \hbox{X-ray} properties of the optical
galaxies selected in \S2.1, so we searched for \hbox{X-ray}
counterparts of these galaxies by cross-correlating the optical
and \hbox{X-ray} source lists with a matching radius of $1\farcs
0$ for sources within $6^{\prime}$ of the average aim point and
$1\farcs 3$ for larger off-axis angles. In 47 cases where there is
more than one optical galaxy associated with an \hbox{X-ray}
source, the closest one is selected. None of the galaxies is
associated with more than one \hbox{X-ray} source. A summary of
these findings is given in \hbox{Table \ref{tblepoch}}. Around one
third of the \hbox{X-ray} sources do not have counterparts in the
galaxy sample, either because their $R$ magnitude is fainter than
25 or they are stars. We also estimated the expected number of
false matches by artificially offsetting the \hbox{X-ray} source
coordinates in right ascension and declination by $5\farcs0$ (both
positive and negative shifts) and re-correlating with the optical
sources. On average, the number of false matches is only $\sim
6\%$ of the total matches in our sample for both the \hbox{CDF-N}
and \hbox{CDF-S}.

\begin{deluxetable}{llccccc}
\tabletypesize{\small} \rotate \tablecaption{Observation Epochs
for the \chandra\ Deep Fields \label{tblepoch}} \tablehead{

&
&
\colhead{Exposure Time}&
\colhead{Obs. Date\tablenotemark{a}} &
\multicolumn{3}{c}{Number of Sources\tablenotemark{b}}  \\

\colhead{Epochs} &
\colhead{Obs. IDs} &
\colhead{(ks)} &
\colhead{$t_{\rm obs}$} &
\colhead{FB} &
\colhead{SB} &
\colhead{HB}  \\
}

\tablewidth{0pt}
\startdata
\hbox{CDF-N} Epoch 1 & 580, 967, 966, 957 & 221.6 & 1999 Dec 12 & 131 & 105 & 73 \\
\hbox{CDF-N} Epoch 2 & 2386, 1671, 2344 & 267.4 & 2000 Nov 22 & 125 & 100 & 76\\
\hbox{CDF-N} Epoch 3 & 2232, 2233, 2423, 2234, 2421 & 488.2 & 2001 Feb 26 & 142 & 118 & 96\\
\hbox{CDF-N} Epoch 4 & 3293, 3388, 3408, 3389 & 385.7 & 2001 Nov 17 & 138  & 115 & 82\\
\hbox{CDF-N} Epoch 5 & 3409, 3294, 3390, 3391 & 581.7 & 2002 Feb 17 & 147 & 128 & 92\\
\hline
\hbox{CDF-S} Epoch 1 & 1431-0, 1431-1 & 114.7 & 1999 Nov 15 & 77 & 68 & 48 \\
\hbox{CDF-S} Epoch 2 & 441, 582 & 186.5 & 2000 Jun 01 & 83 & 68 & 53\\
\hbox{CDF-S} Epoch 3 & 2406, 2405, 2312, 1672, 2409, 2313, 2239 & 638.2 & 2000 Dec 18 & 136 & 124 & 81\\
\hbox{CDF-S} Epoch 4 F01\tablenotemark{c} & 5015, 5016 & 240.1 & 2004 Mar 01 & 52 & 44 & 39\\
\hbox{CDF-S} Epoch 4 F02\tablenotemark{c} & 5017, 5018 & 227.4 & 2004 May 15 & 40 & 36 & 29\\
\hbox{CDF-S} Epoch 4 F03\tablenotemark{c} & 5019, 5020 & 240.7 & 2004 Nov 17 & 22 & 18 & 13\\
\hbox{CDF-S} Epoch 4 F04\tablenotemark{c} & 5021, 5022, 6164 & 246.0 & 2004 Nov 16 & 26 & 24 & 18\\
\enddata

\tablenotetext{a}{Observations within each epoch only lasted for a
few days, so we neglected the length of the epochs and only show
the average observation dates weighted by exposure time.}
\tablenotetext{b}{Number of galaxies in the sample that are
\hbox{X-ray} detected in this epoch for the particular band:
\hbox{0.5--8.0~keV} (FB),  \hbox{0.5--2.0~keV} (SB), or
\hbox{2--8~keV} (HB).} \tablenotetext{c}{Observational fields of
the E-CDF-S. Note that a large fraction of each of these fields
extends outside the region of our outburst survey.}

\end{deluxetable}

\subsection{Source Characterization and Outburst Searching}

We measured the \hbox{X-ray} counts coincident with every optical
galaxy in all 36 images using circular-aperture photometry. We
have chosen the aperture radii based on the encircled-energy
function of the \chandra\ PSF, which was calculated using MARX
(Model of AXAF Response to \hbox{X-rays}\footnote{See
http://space.mit.edu/CXC/MARX}) simulations. The 90\%
encircled-energy radius of the PSF was used. The circular aperture
was centered at the position of every optical galaxy or its
\hbox{X-ray} counterpart if it was \hbox{X-ray} detected. If a
galaxy was not fully covered by a certain image, then it was
marked as undetectable for this image.

The local background was determined in an annulus outside of the
source-extraction region using background maps with known
\hbox{X-ray} sources carefully removed to avoid possible
contamination from adjacent sources. To create a background map
for a given image, we first merged our \chandra\ \hbox{X-ray}
source list from this image to the main {\it Chandra} catalogs
\citep{Alexander2003} to get a combined source list, using a
matching radius of $2\farcs 5 $ for sources within $6^{\prime}$ of
the average aim point and $4\farcs 0$ for larger off-axis angles.
Then we masked out these sources using apertures with radii twice
that of the 90\% PSF encircled-energy radius. We filled the masked
regions for each source with a local background estimated by
making a probability distribution of counts using an annulus with
inner and outer radii of 2 and 4 times the 90\% PSF
encircled-energy radius, respectively. The local background for
every optical galaxy was then determined using the same annulus on
this background map.

When an optical galaxy was \hbox{X-ray} detected (i.e., had an
\hbox{X-ray} counterpart) in a given image, the net number of
source counts was calculated by subtracting the expected number of
background counts from the number of counts in the aperture.
Poisson errors were calculated following \citet{Gehrels1986} and
were propagated through this calculation. When an optical galaxy
was not detected, an upper limit was calculated. If the number of
counts in the aperture was $\le10$, the upper limit was calculated
using the Bayesian method of \citet*{Kraft1991} for 99.87\%
confidence. For a larger number of counts in the aperture, a
$3\sigma$ upper limit was set by multiplying the square root of
the number of background counts by three. The number of source
counts was then divided by the effective exposure time, which was
the average value within the aperture on the exposure map, to get
the \hbox{X-ray} count rate (this procedure corrects the count
rate for vignetting and other effects).

The E-CDF-S, which serves as the fourth epoch for the
\hbox{CDF-S}, contains four observational fields, which overlap in
a few areas over $\sim$50 ${\rm arcmin}^2$. If an optical galaxy
was detected in more than one field, we chose data from the field
with greater source counts. If it was detected in only one field,
we chose data from this field. If it was not detected in any field
but was in the overlapping area, we chose data from the field with
the longest effective exposure time.

We searched for outbursts following these steps:

\begin{enumerate}

\item For every optical galaxy, we calculated its count rate or
upper limit on count rate in each epoch and each energy band
(i.e., each of the 36 images), unless it was not fully covered by
a given image. A galaxy had coverage in up to five epochs if it
was in the \hbox{CDF-N} and up to four epochs if it was in the
\hbox{CDF-S}.

\item For each energy band, we selected galaxies that were
detected in at least one epoch. For each of these galaxies, we
compared its highest count rate with its lowest count rate or
upper limit on count rate. Thus we got either the variability
factor or its lower limit for each source. If there was a
variation of more than a factor of 20, then we considered that
there was a candidate for an \hbox{X-ray} outburst in this galaxy.

\end{enumerate}

\section{RESULTS AND DISCUSSIONS}
\begin{figure}
\centerline{\includegraphics[scale=0.5]{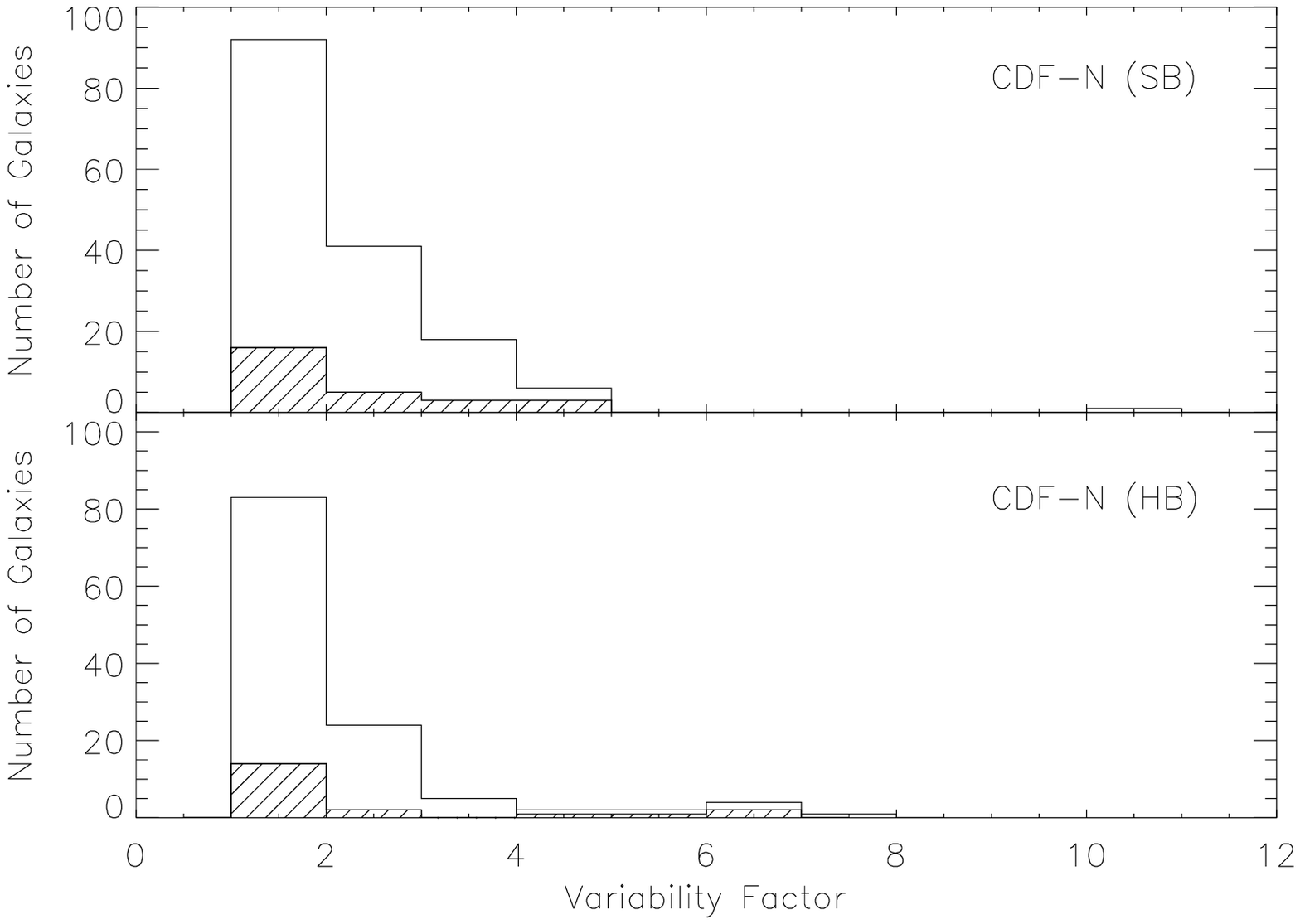}
\includegraphics[scale=0.5]{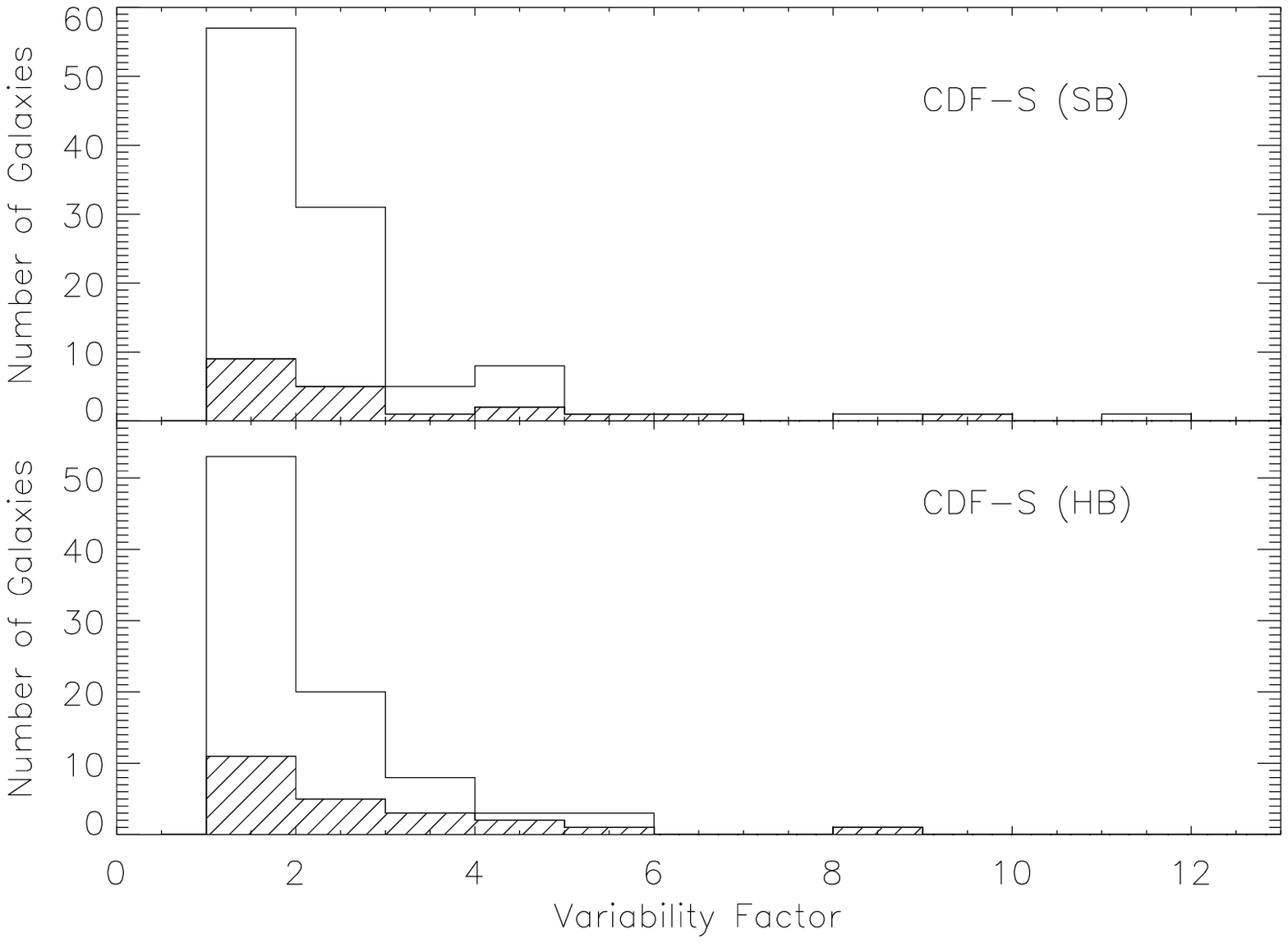}
} \figcaption{Histograms showing the distribution of variability
factors in the SB (upper panels) or HB (lower panels) for galaxies
in the \hbox{CDF-N} and \hbox{CDF-S}. The variability factor of a
given galaxy is defined as the ratio between its highest count
rate and its lowest count rate or upper limit on count rate. The
shaded areas are for galaxies with lower limits on variability
factors only. \label{vari}}
\end{figure}

After systematically analyzing the count-rate variations, we found
no outbursts in either the \hbox{CDF-N} or in the \hbox{CDF-S}.
The distribution of the variability factors in the SB and HB is
shown in Figure \ref{vari}. The median relative uncertainty of
these factors is $\sim 30\%$. The count-rate variations of a few
sources exceed a factor of 10. However, these are all off-axis
sources and the significances of the variations are $<1 \sigma$.

\subsection{Upper Limits on the Outburst Event Rate}

\begin{figure}
\centerline{\includegraphics[scale=0.5]{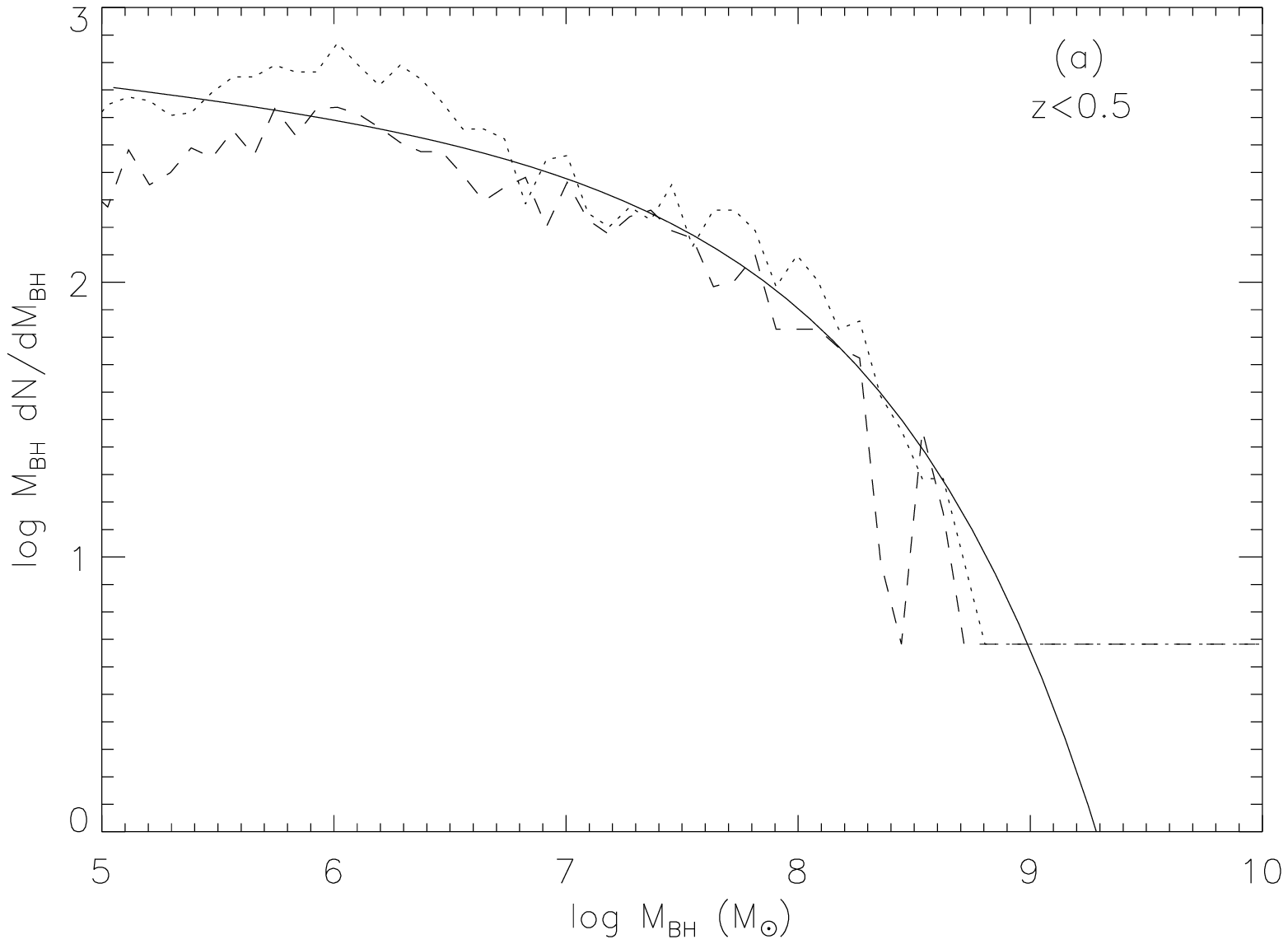}
\includegraphics[scale=0.5]{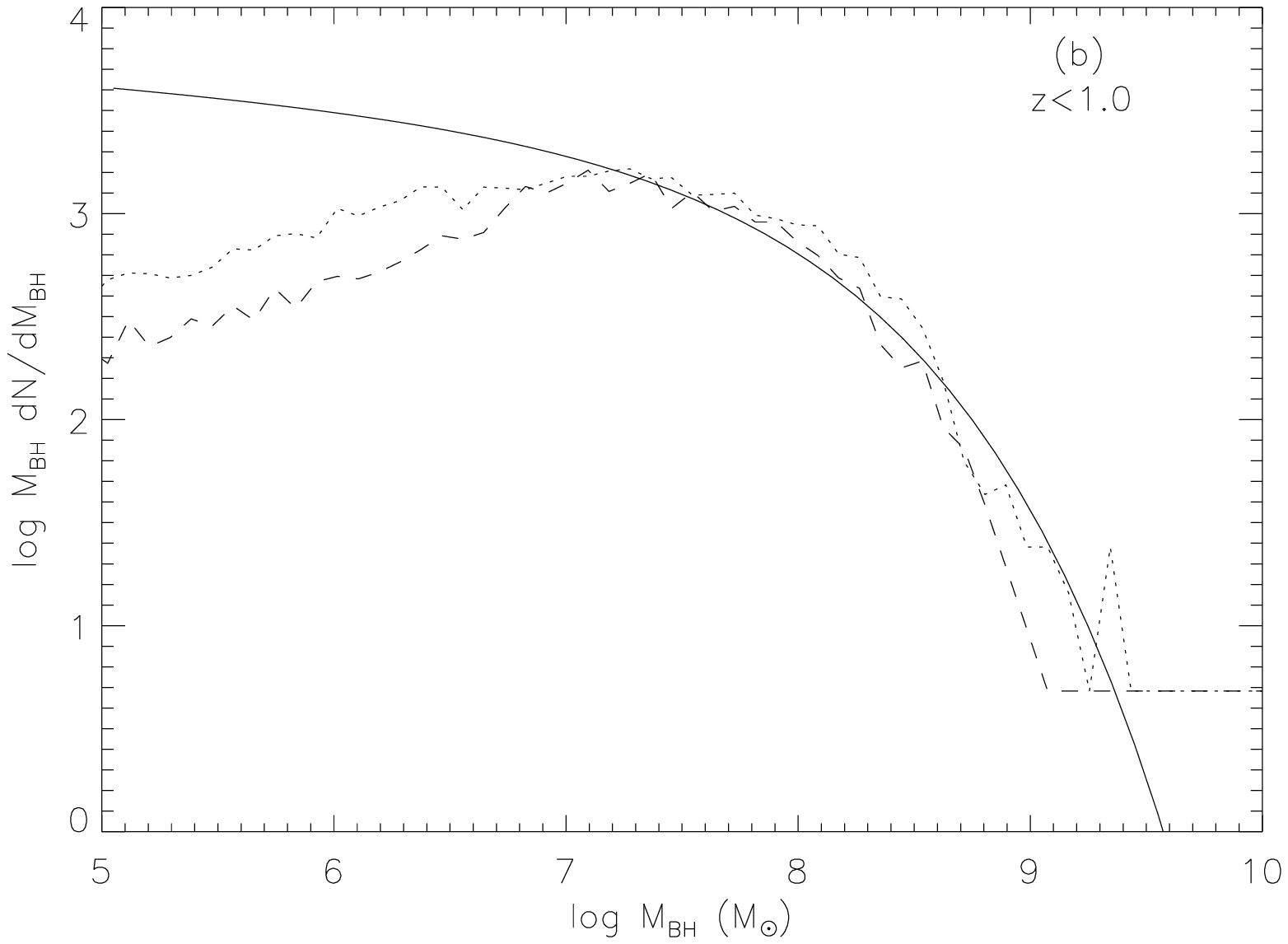}
} \figcaption{SMBH mass functions for galaxies with redshift less
than ({\it a}) 0.5 or ({\it b}) 1.0, and the local SMBH mass
function from \citet[][solid curve]{Marconi2004} scaled to compare
with the mass functions for our sample. The dotted curves indicate
mass functions in the \hbox{CDF-N} while the dashed curves are for
the \hbox{CDF-S}. The deviation at low SMBH mass in the right
panel is simply due to a selection effect; it is difficult to
detect the galaxies hosting SMBHs with small masses at relatively
high redshifts. \label{bhmf}}
\end{figure}

The nondetection of any \hbox{X-ray} outbursts in this survey can
be used to constrain the rate of such outbursts in the Universe.
Since the detectability of an outburst depends upon its
luminosity, the constraints will have a luminosity dependence.
Moreover, not all galaxies in this survey are capable of producing
outbursts. Galaxies without a central SMBH (e.g., dwarf
irregulars) cannot tidally disrupt stars, and a SMBH of mass
$M_{\rm BH} \gtrsim 3 \times 10^8M_{\sun}$ will swallow stars
whole \citep[e.g.,][]{Frank1976}. SMBH candidates with masses of a
few $10^5M_{\sun}$ in the centers of galaxies have been found
\citep[e.g.,] []{Greene2004,Peterson2005}. Thus we consider that
only galaxies with a SMBH mass greater than $10^5M_{\sun}$ and
less than $3 \times 10^8M_{\sun}$ can produce outbursts. SMBH
masses were roughly estimated for all galaxies in this survey with
the relation between $M_{\rm BH}$ and total galaxy luminosity in
the $K$ band, $L_{\rm K, total}$. The relation was derived from
the data for 27 low-redshift galaxies (10 late-type and 17
early-type galaxies) in \citet{Marconi2003}. The intrinsic
dispersion of this relation is $\sim$0.5 dex in \hbox{log $M_{\rm
BH}$}. Although there is evidence showing cosmic evolution of the
$M_{\rm BH}$-$\sigma$ (SMBH mass and bulge velocity dispersion)
relation and the bulge-to-SMBH mass ratio
\citep[e.g.,][]{Woo2006}, the coexistent luminosity evolution
makes the SMBHs at $z\gtrsim1$ coincidently fall on nearly the
same $M_{\rm BH}$ versus $R$-band magnitude ($M_{\rm R}$) relation
(to 0.3 mag) as low-redshift galaxies \citep{Peng2006}. Thus we
can expect that the $M_{\rm BH}$-$L_{\rm K, total}$ relation also
holds approximately at $z\sim 0.8$ for this galaxy sample. $L_{\rm
K, total}$ for the galaxies in this survey was extrapolated from
$HK^{\prime}$-band magnitudes \citep[10826 galaxies,][]{Capak2004}
or $z^{\prime}$-band magnitudes \citep[2873
galaxies,][]{Capak2004} for the \hbox{CDF-N}, and from $z$-band
magnitudes \citep[10167 galaxies,][]{Caldwell2005} or $R$-band
magnitudes \citep[802 galaxies,][]{Wolf2004} for the \hbox{CDF-S},
using the spectral energy distribution for Sbc galaxies
\citep*{Coleman1980}. The resulting conversions from
$HK^{\prime}$, $z^{\prime}$, $z$ and $R$ magnitudes (AB system) to
the standard $K$ magnitude were given by $HK^{\prime}-K=1.65$,
$z^{\prime}-K=2.07$, $z-K=2.01$, and $R-K=2.70$. Based on the
dispersion of the $M_{\rm BH}$-$L_{\rm K, total}$ relation and the
uncertainties in the color conversions, the derived $M_{\rm BH}$
is estimated to be good within an order of magnitude (as we show
below, our main results are not sensitive to $M_{\rm BH}$). A
comparison between the SMBH mass functions derived for the
\hbox{low-redshift} galaxies in this sample and the local SMBH
mass function for all galaxy types \citep{Marconi2004} is shown in
Figure \ref{bhmf}. There is basic agreement between the shapes of
these mass functions indicating that our mass measurements are
reasonable.

There are $11\,339$ galaxies in the \hbox{CDF-N} and $8\,268$
galaxies in the \hbox{CDF-S} with $M_{\rm BH}$ in the range of
$10^5M_{\sun} < M_{\rm BH} < 3 \times 10^8M_{\sun}$. There will
also be another constraint set by the physics of outburst
production, i.e., a SMBH with a given mass cannot produce
outbursts with arbitrarily high luminosities owing to, e.g., the
Eddington limit. However, we ignore this constraint for now, as we
would first like to present model-independent limits.

\begin{figure}
\centerline{\includegraphics[scale=0.5]{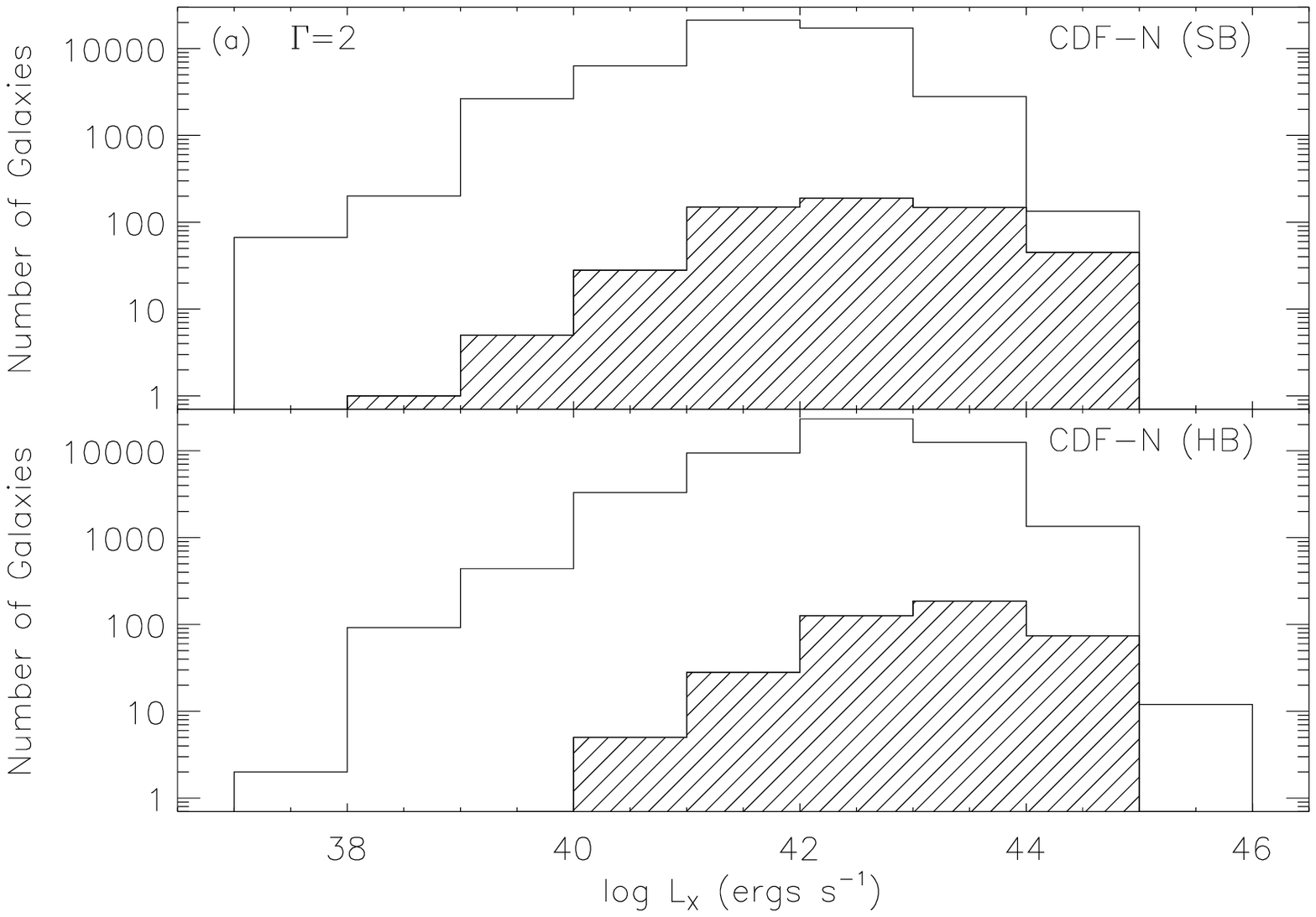}
\includegraphics[scale=0.5]{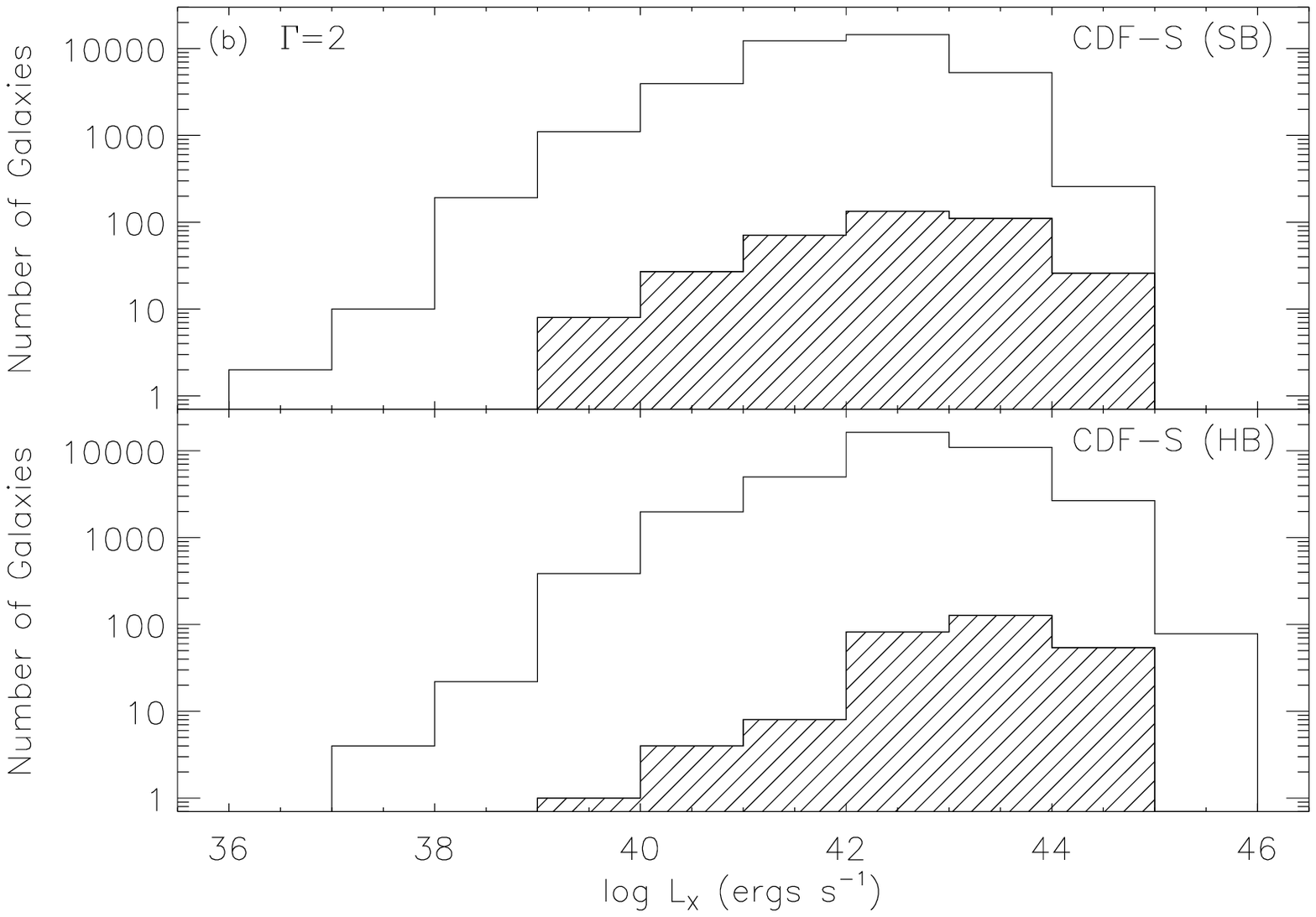}
}

\centerline{\includegraphics[scale=0.5]{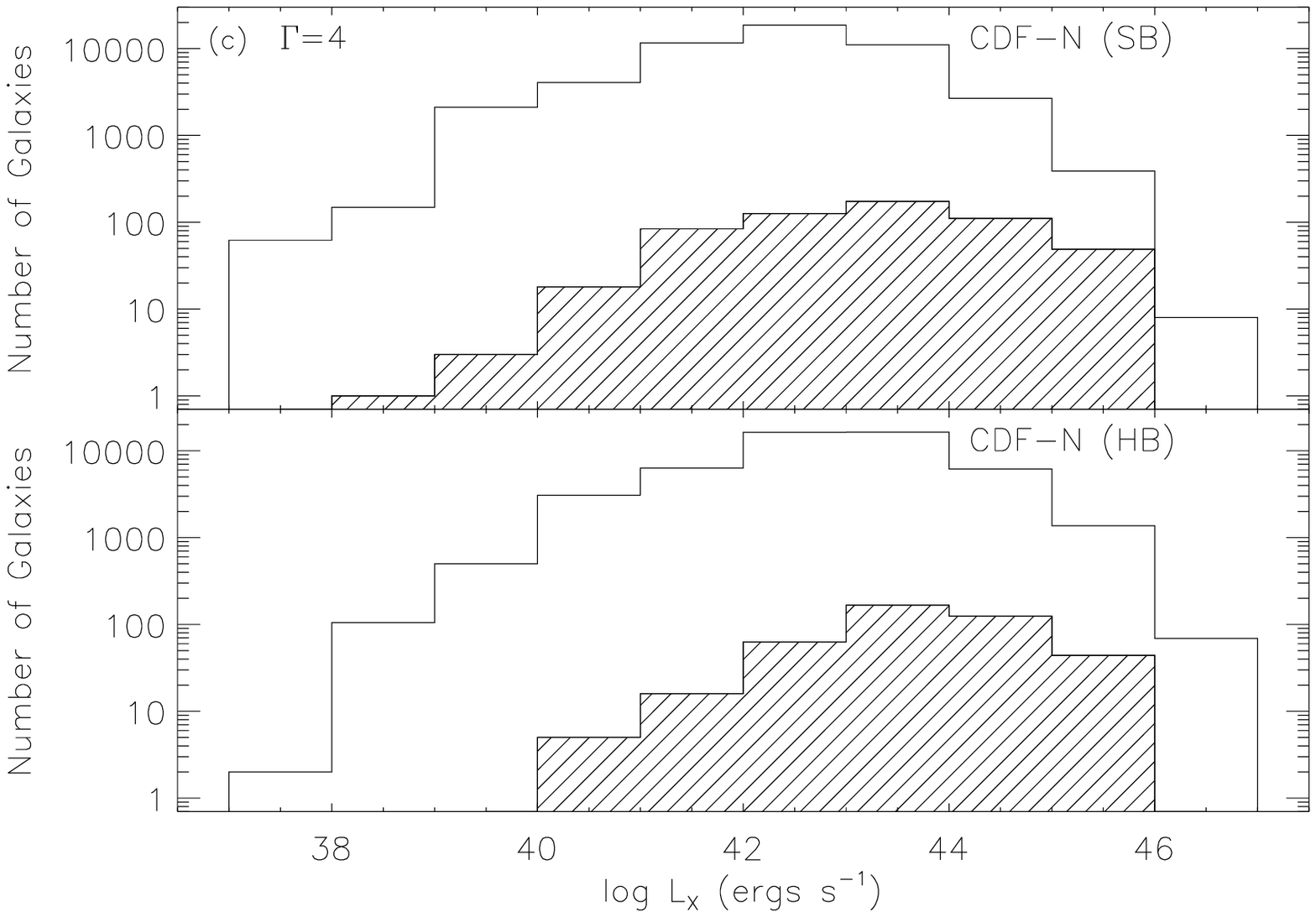}
\includegraphics[scale=0.5]{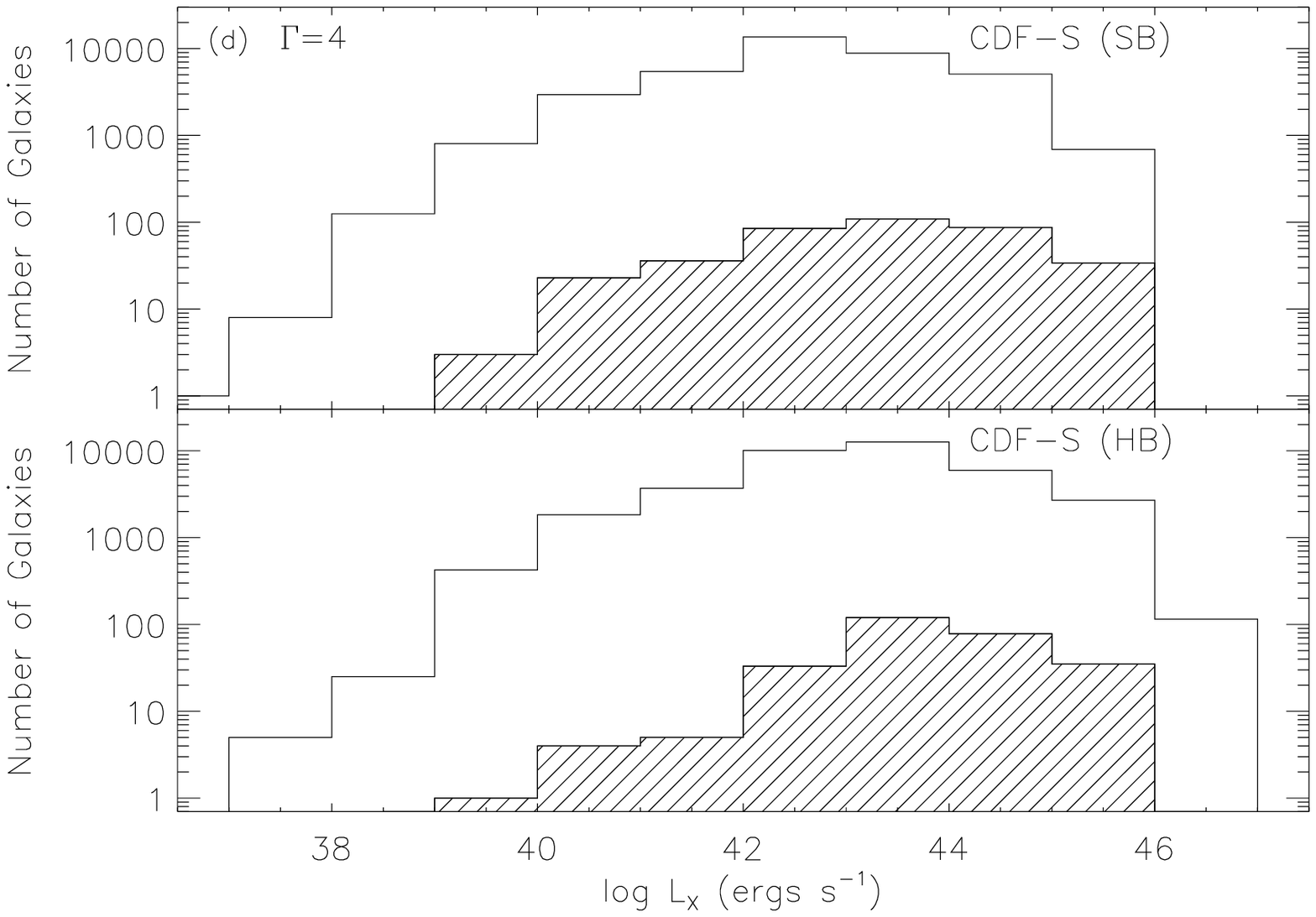}
}

\figcaption{SB and HB \hbox{X-ray} luminosity distributions of
galaxies in the \hbox{CDF-N} and \hbox{CDF-S}. Luminosities for
all the epochs were collected, so each galaxy is plotted multiple
times, corresponding to the number of epochs in which it was
observed. Most of the luminosities are $3\sigma$ upper limits.
Luminosities of \hbox{X-ray} detected galaxies are represented by
the shaded area. The number of galaxies was plotted using a
logarithmic scale to show the small fraction of \hbox{X-ray}
detected galaxies. A photon index of $\Gamma=2$ or $\Gamma=4$ was
adopted in these plots. \label{ldist}}
\end{figure}

To determine the allowed rate of outbursts of \hbox{X-ray}
luminosity greater than or equal to a certain value, we used the
following ``recipe'':
\begin{enumerate}

\item For a given energy band (FB, SB, or HB), we picked an
\hbox{X-ray} luminosity, $L_{\rm X, burst}$, as the lower limit,
and we derived the rate of outbursts of \hbox{X-ray} luminosities
$\ga L_{\rm X, burst}$. Count rates or upper limits on count rates
for all the galaxies in the survey were converted to luminosities
based on their redshifts, using the Portable, Interactive,
Multi-Mission Simulator (PIMMS) for a power-law photon index of
$\Gamma$. Galaxies generally have soft \hbox{X-ray} spectra during
outburst and harder spectra in the their basal states. We thus
considered a few choices of $\Gamma$ ranging from 2 to 5. The
luminosity distributions for $\Gamma=2$ and $\Gamma=4$ are shown
in Figure \ref{ldist}. Most of the luminosities are upper limits.

\item We assumed that the typical duration of an outburst was
$T_{\rm burst}$. Theoretically predicted light curves of outbursts
show a characteristic fast rise and slow decay. After the bulk of
the material is accreted, which could be on a timescale of the
order of months, the debris starts to form a radiation-pressure
supported torus and the luminosity declines slowly as $\propto
t^{-5/3}$ \citep{Rees1988,Rees1990}. This long-term decay could
last for years and may have been observed (with large light-curve
gaps) in a few outbursts \citep[e.g.,][]{Komossa99}. The detailed
luminosity evolution of outbursts, especially for the first few
months, remains unknown, and likely varies greatly from
event-to-event. For simplicity, we assumed that the luminosity is
constant during the outburst, and after that the \hbox{X-ray}
luminosity would drop by a minimum factor of 20. We then
determined the total rest-frame time over which we are sensitive
to outbursts of luminosity $\ga L_{\rm X, burst}$:
\begin{equation}
T_{\rm total}(L_{\rm X, burst},T_{\rm burst})=\sum_{i=1}^{n}
{T_{i, {\rm sens}}(L_{\rm X, burst}, T_{\rm burst})},
\label{eq_ttotal}
\end{equation}
\noindent where $n=19607$ is the number of galaxies in the survey
with \hbox{$10^5M_{\sun} < M_{\rm BH} < 3 \times 10^8M_{\sun}$},
and $T_{i, {\rm sens}}$ is the rest-frame time over which we are
sensitive to an outburst of luminosity $\ga L_{\rm X, burst}$ for
galaxy $i$. $T_{i, {\rm sens}}$ can be expressed as:
\begin{equation}
T_{i, {\rm sens}}=\int_{\rm Time,rf}{W_i(t) dt}.
\end{equation}
\noindent This integration is over all the rest-frame observation
period, from time $T_{\rm burst}$ before the first epoch to the
last epoch, and $W_i(t)$ is our sensitivity ``window function'' to
outbursts starting from time $t$ and with luminosity $\ga L_{\rm
X, burst}$ and duration $T_{\rm burst}$. If we could detect such
an outburst based on our searching criteria (step~2 in \S2.3),
then $W_i(t)=1$; otherwise $W_i(t)=0$.
%

\item The 90\% confidence upper limit on 0 events is 2.303 from
\citet{Gehrels1986}. Thus the 90\% confidence upper limit on the
event rate of outbursts is given by
\begin{equation}
\dot{N}_{\rm CDF}(L_{\rm X, burst},T_{\rm burst})=\frac{2.303}{T_{\rm total}} \ \ \
\frac{\rm outbursts}{\rm galaxy\ yr}.
\label{ncdf}
\end{equation}

\item Repeat the above steps for various values of $L_{\rm X,
burst}$ and $T_{\rm burst}$ as well as different assumptions about
outburst spectral shape.
\end{enumerate}

\begin{figure}
\centerline{\includegraphics[scale=0.5]{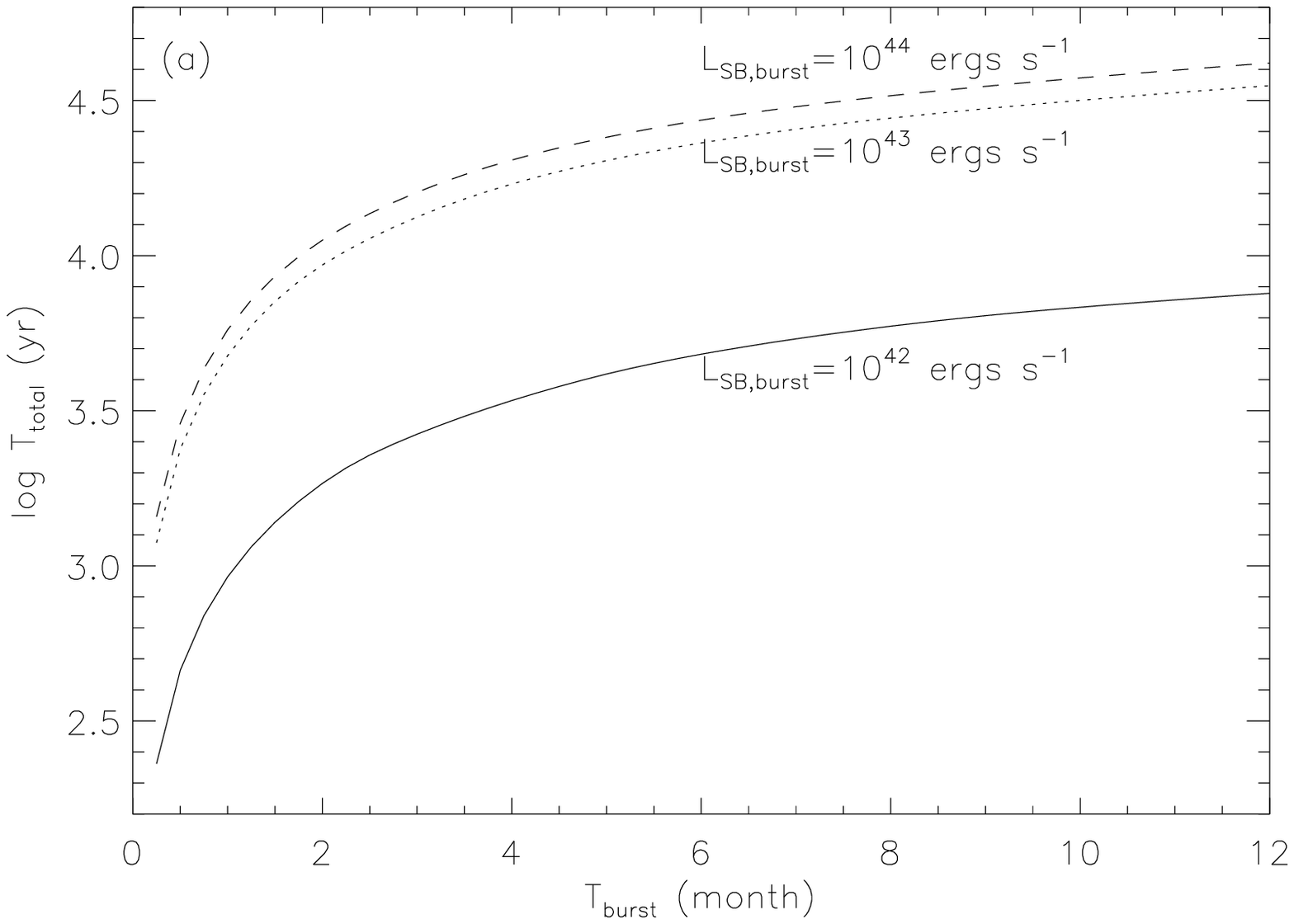}
\includegraphics[scale=0.5]{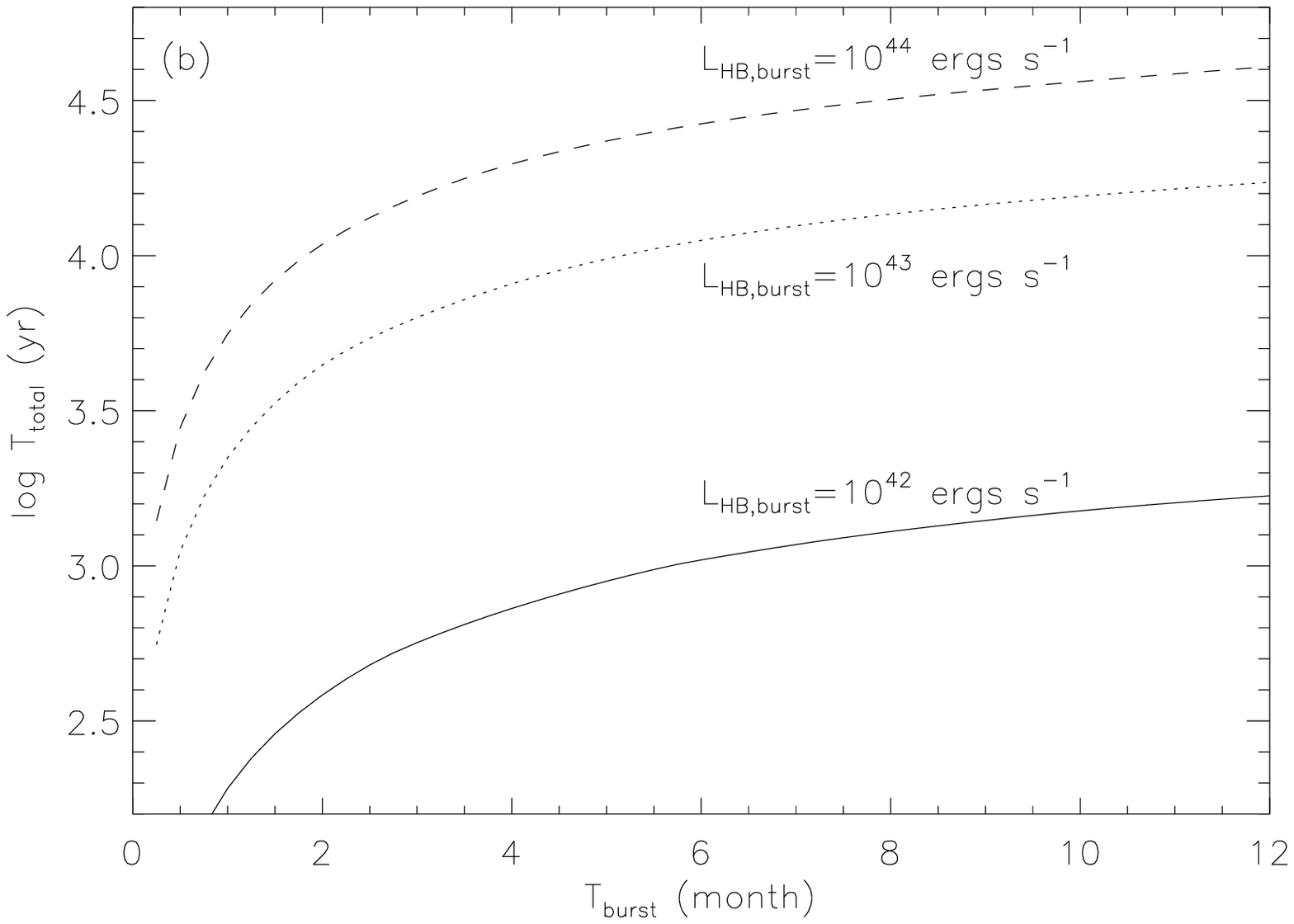}
} \figcaption{Dependence of $T_{\rm total}$ on outburst duration
at different outburst luminosities. $T_{\rm total}$ is the total
rest-frame time over which we are sensitive to outbursts of ({\it
a}) SB luminosity $L_{\rm SB, burst}$ or ({\it b}) HB luminosity
$L_{\rm HB, burst}$; see \hbox{ Equation \ref{eq_ttotal}}. A
photon index of $\Gamma=2$ was adopted when making these plots.
\label{ttotal2}}
\end{figure}

\begin{figure}
\centerline{\includegraphics[scale=0.5]{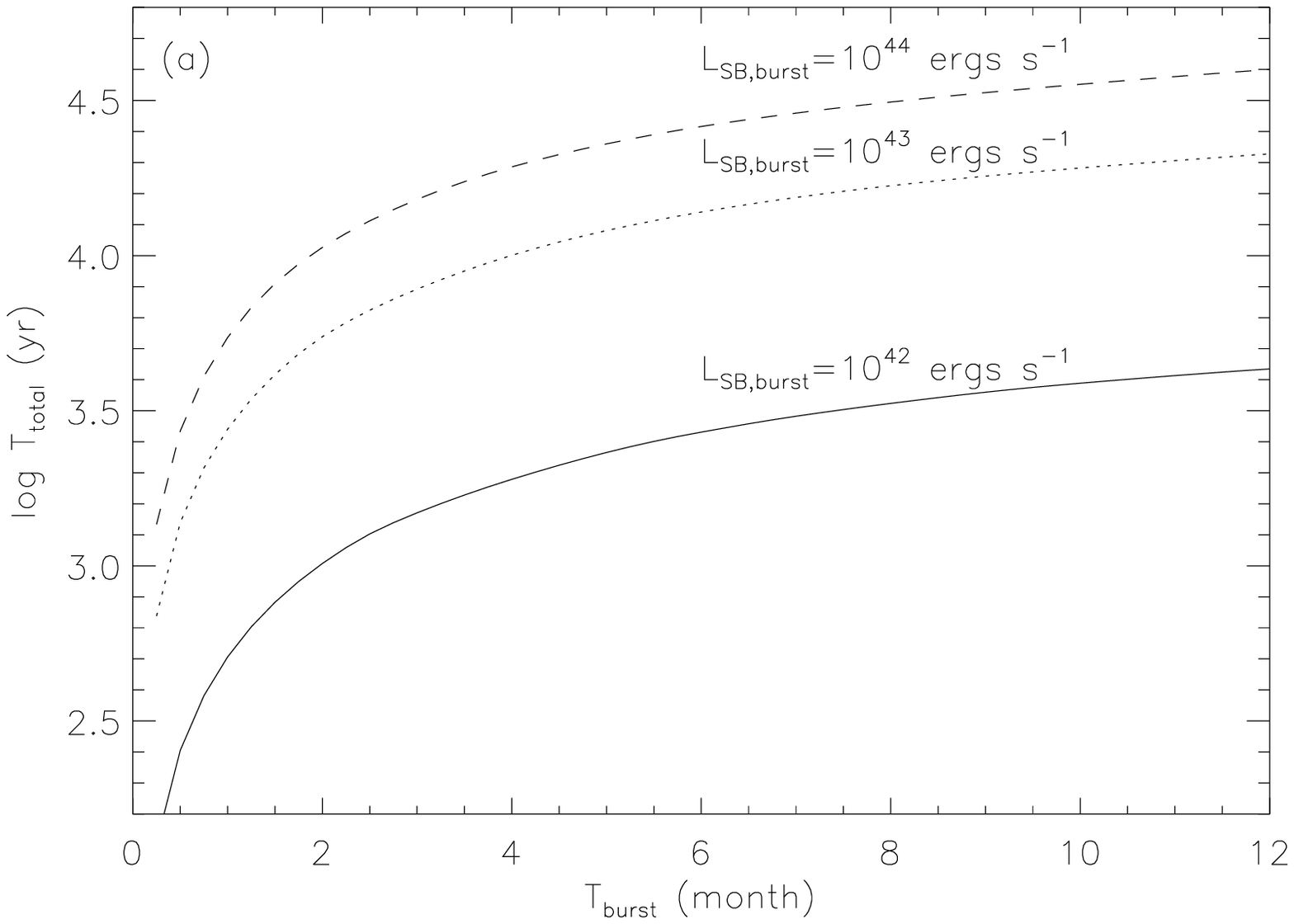}
\includegraphics[scale=0.5]{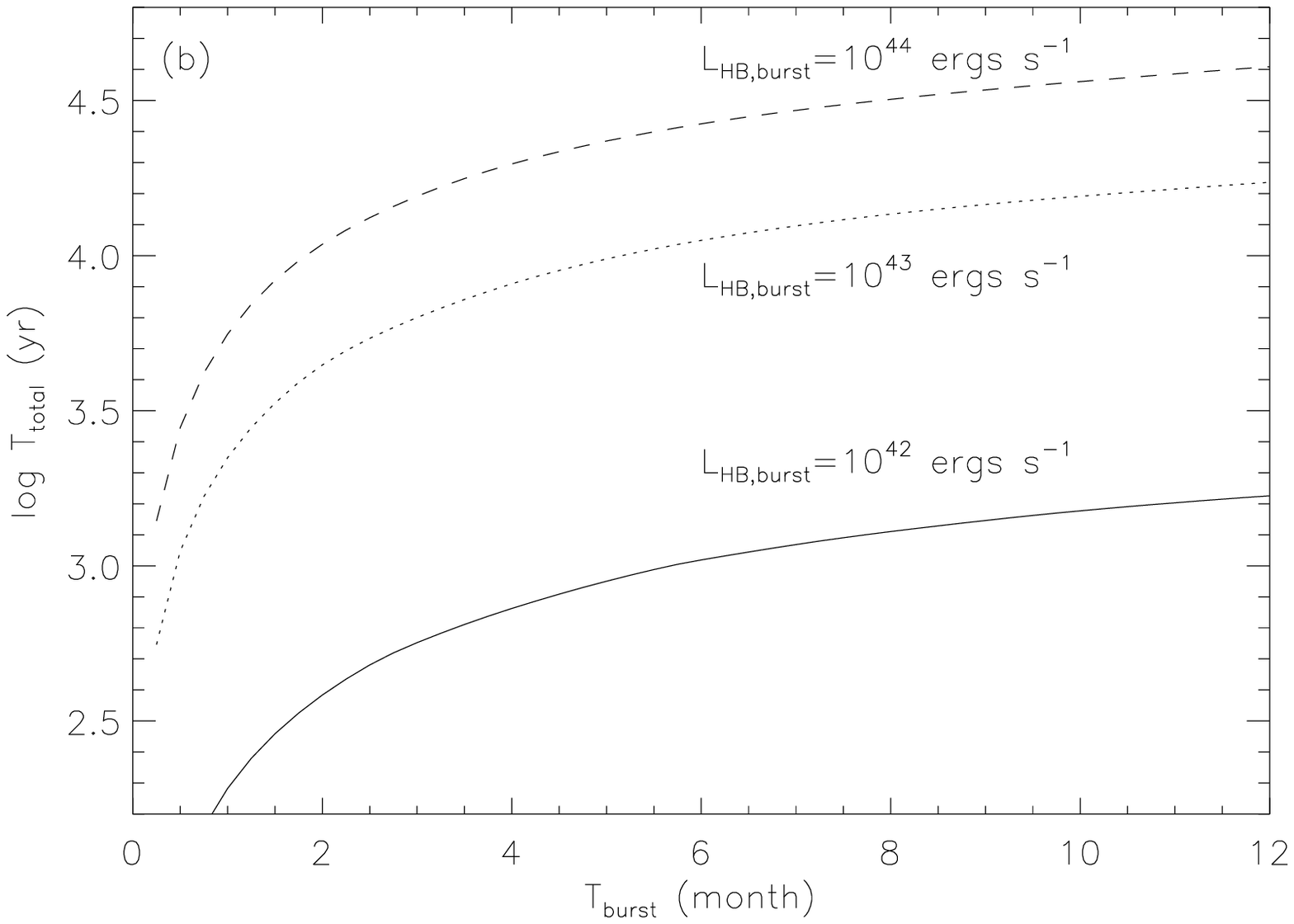}
} \figcaption{The same as Figure \ref{ttotal2}, but for a photon
index of $\Gamma=4$. \label{ttotal4}}
\end{figure}

\begin{figure}
\centerline{\includegraphics[scale=0.5]{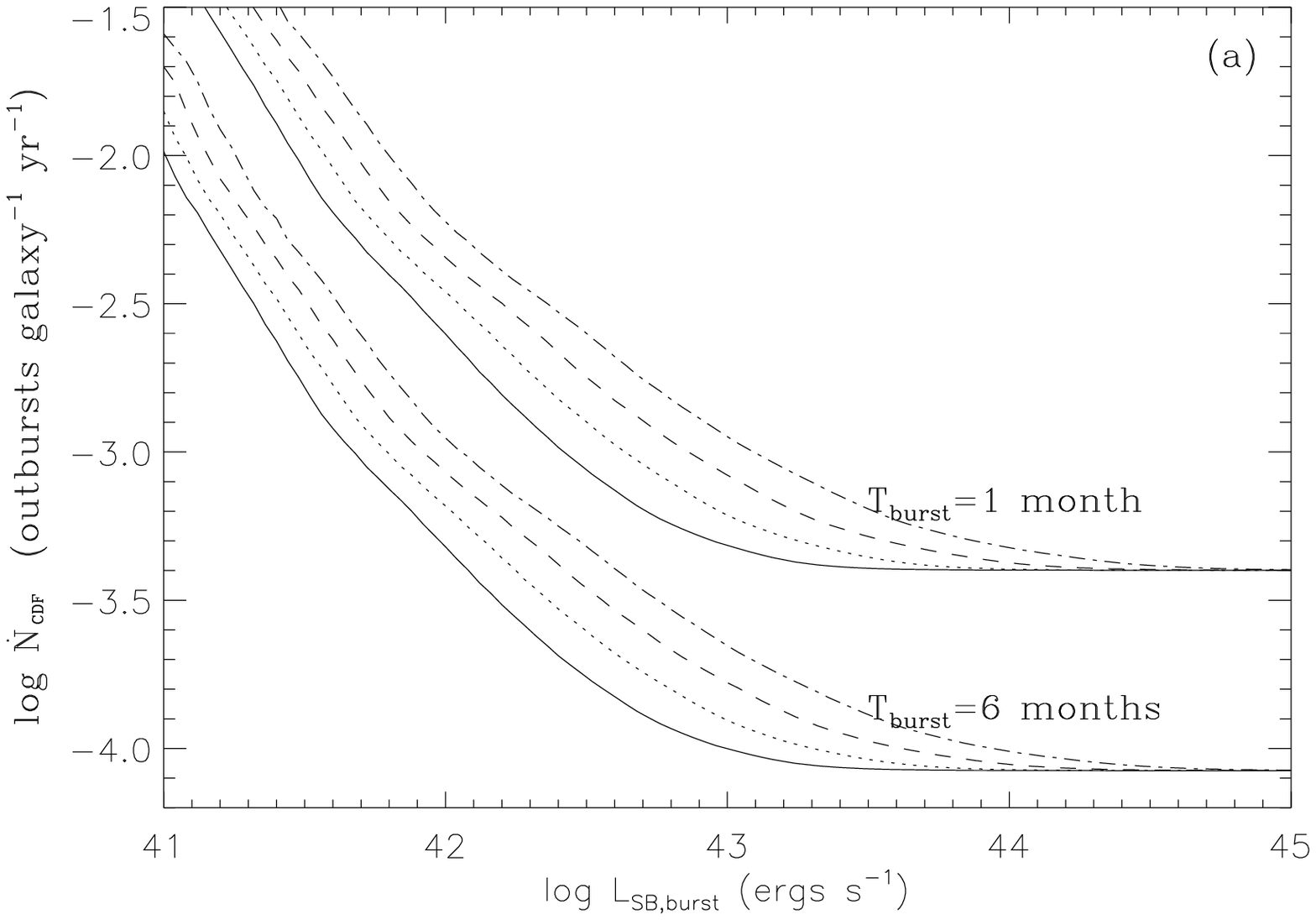}
\includegraphics[scale=0.5]{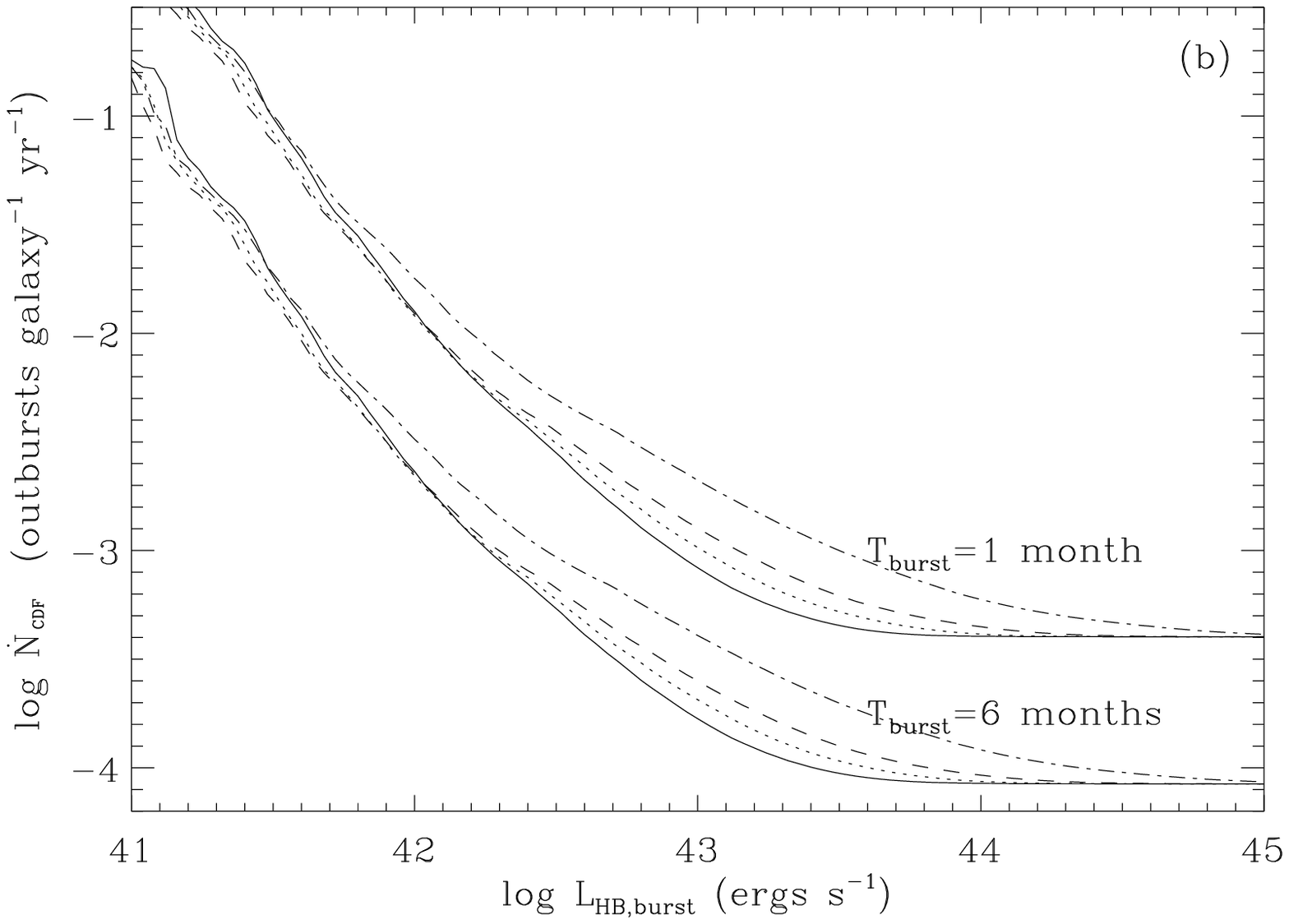}
} \figcaption{90\% confidence upper limits on the event rates of
outbursts derived from this survey as functions of the lower limit
on ({\it a}) SB or ({\it b}) HB \hbox{X-ray} luminosity; see
Equation \ref{ncdf}. Solid, dotted, dashed, and dash-dotted lines
represent photon indices $\Gamma$ of 2, 3, 4, and 5, respectively.
We show the constraints on two different outburst durations, 1
month and 6 months. We made no assumptions about the physical
process causing the outburst except that the SMBH mass is in the
range $10^5M_{\sun} < M_{\rm BH} < 3 \times 10^8M_{\sun}$.
\label{ever}}
\end{figure}

The constraints set by Equation \ref{ncdf} depend significantly on
the outburst duration $T_{\rm burst}$ and the lower luminosity
limit $L_{\rm X, burst}$, and they weakly depend on the spectral
shape (we consider \hbox{X-ray} photon indices $\Gamma=$2, 3, 4,
or 5) and energy band (the SB or HB), as shown in Figures
\ref{ttotal2}, \ref{ttotal4} and \ref{ever}. Longer outburst
durations lead to tighter upper limits on the event rate. Photon
indices affect the derived \hbox{X-ray} luminosities through the
conversion from count rate to flux (from PIMMS) and the $K$
correction term $(1+z)^{(\Gamma-2)}$. As $\Gamma$ increases, the
conversion factor increases in the SB and decreases in the HB,
while the $K$ correction term always increases and preferentially
dominates over the conversion factor for high-luminosity sources
since they generally have large redshifts. Thus the dependence of
the upper limits on the spectral shape behaves differently in the
SB and HB. Note that luminosities $L_{\rm X, burst}$ in the SB and
HB are different, so the upper limits on the event rate in these
energy bands cannot be compared to each other directly. Generally,
assuming an outburst duration of 6~months, which was also adopted
in Donley et~al. (2002), the upper limit on the event rate is
$\sim 10^{-4}$~galaxy$^{-1}$~yr$^{-1}$, for an outburst with
\hbox{X-ray} luminosity $\ga 10^{43}$~ergs~${\rm s}^{-1}$.

\subsection{Comparison with Previous Results}

We compared these new constraints with the theoretical study by
\citet{Wang2004} using a singular isothermal sphere, which
analytically predicted a stellar tidal-disruption rate of
\begin{equation}
\dot{N}_{\rm WM}(M_{\rm BH})\approx 7.1 \times 10^{-4}\ {\rm
yr}^{-1} \left(\frac{\sigma}{70~{\rm
km~s}^{-1}}\right)^{7/2}\left(\frac{M_{\rm BH}} {10^6M_{\sun}}
\right)^{-1}\left(\frac{m_{\star}}{M_{\sun}} \right)^{-1/3}
\left(\frac{R_{\star}}{R_{\sun}} \right)^{1/4}A(z), \label{nth0}
\end{equation}
\noindent with $M_{\rm BH}$ being the SMBH mass, $\sigma$ the
velocity dispersion of the host galaxy, and $m_{\star}$ and
$R_{\star}$ the mass and radius of the tidally disrupted stars. We
added an ``amplification'' factor $A(z)$ here to represent any
redshift evolution of the rate. As the evolution of \hbox{X-ray}
outbursts is unknown (see \S1), we set $A(z)=1$ for now. Utilizing
the $M_{\rm BH}$-$\sigma$ relation from \citet{Ferrarese2005}:
\begin{equation}
M_{\rm BH}=1.66\times 10^8 M_{\sun}\left(\frac{\sigma}{200~{\rm
km~s}^{-1}} \right)^{4.86},\label{msig}
\end{equation}
\noindent Equation \ref{nth0} becomes
\begin{equation}
\dot{N}_{\rm WM}(M_{\rm BH})\approx 7.0 \times 10^{-4}\ {\rm yr}^{-1}
\left(\frac{M_{\rm BH}}
{10^6M_{\sun}} \right)^{-0.28}\left(\frac{m_{\star}}{M_{\sun}} \right)^{-1/3}
\left(\frac{R_{\star}}{R_{\sun}} \right)^{1/4}A(z).
\label{nth}
\end{equation}

\begin{figure}
\centerline{\includegraphics[scale=0.5]{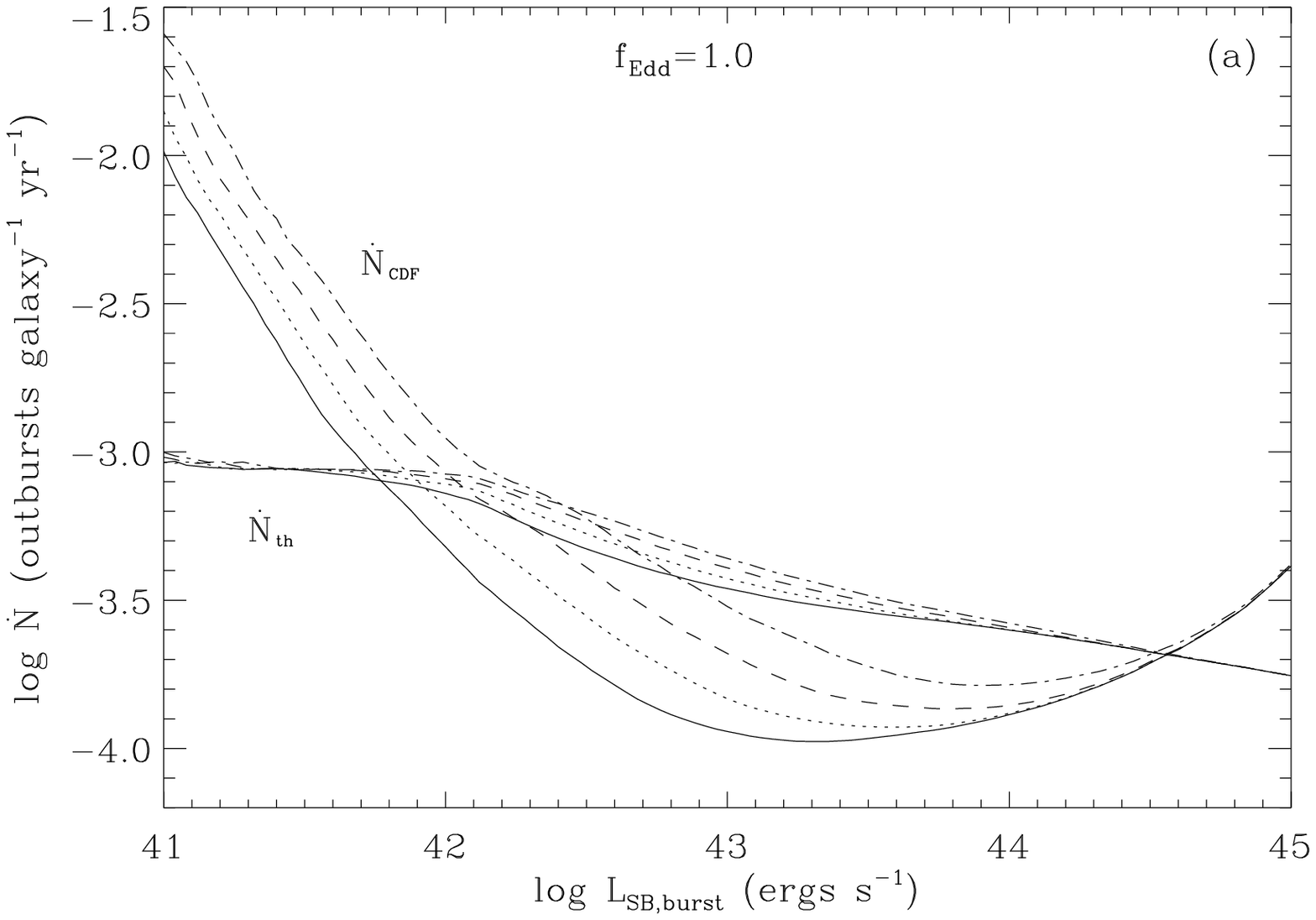}
\includegraphics[scale=0.5]{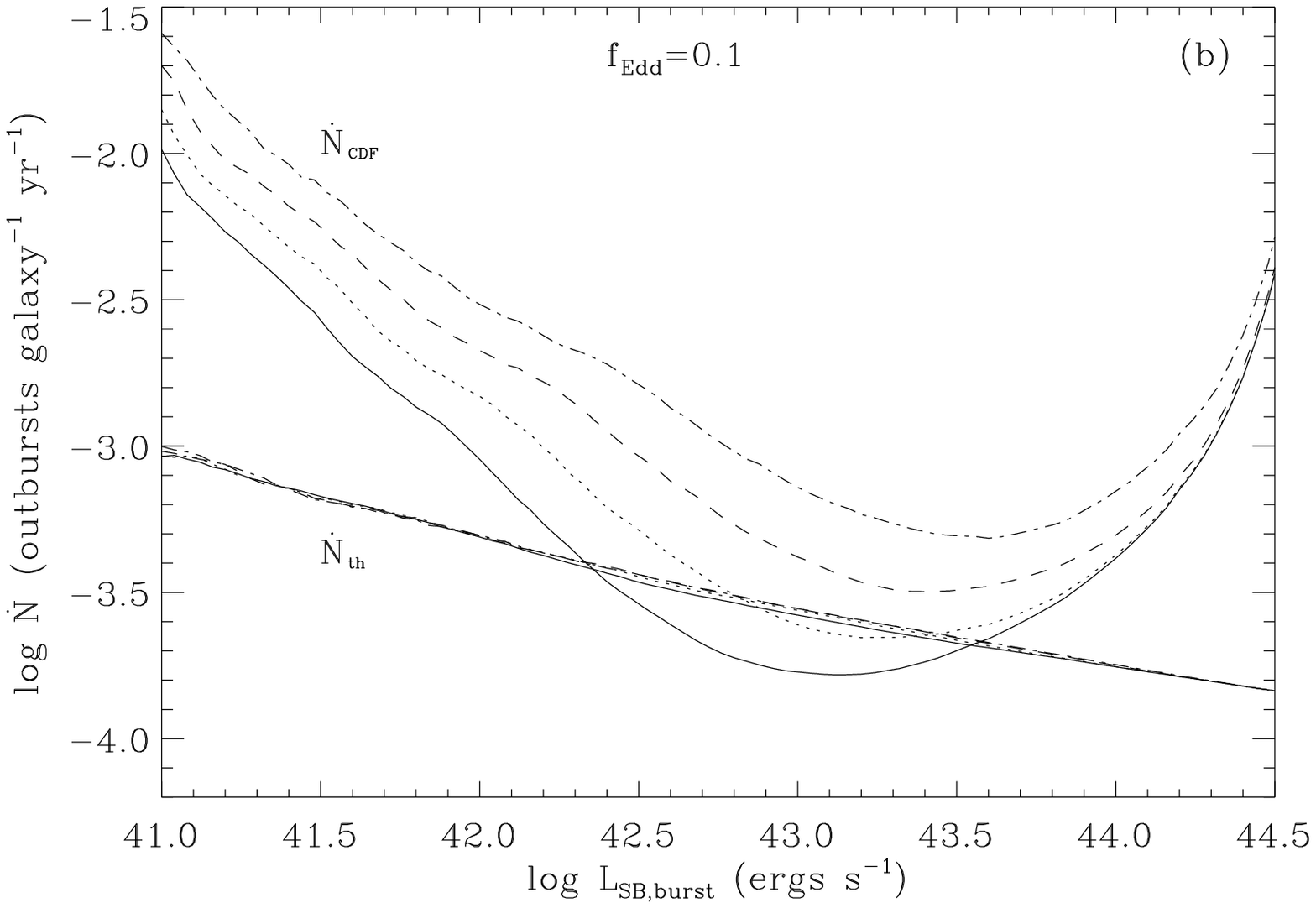}}
\figcaption{90\% confidence upper limits on the event rates of
outbursts derived from this survey and expected event rates from
the theoretical prediction as functions of the lower limit on the
SB \hbox{X-ray} luminosity of outbursts. Solid, dotted, dashed,
and dash-dotted lines represent photon indices $\Gamma$ of 2, 3,
4, and 5, respectively. SMBHs that were not capable of producing
outbursts of luminosity $L_{\rm SB, burst}$  were removed from the
sample. We assumed $T_{\rm burst}=6$ months, $f_{\rm bc}=10$ and
$f_{\rm Edd}=1.0$ ({\it a}) or 0.1 ({\it b}). The predicted rates
were derived under the assumption of isothermal stellar density
distributions in galactic nuclei. \label{everth}}
\end{figure}

\begin{figure}
\centerline{\includegraphics[scale=0.5]{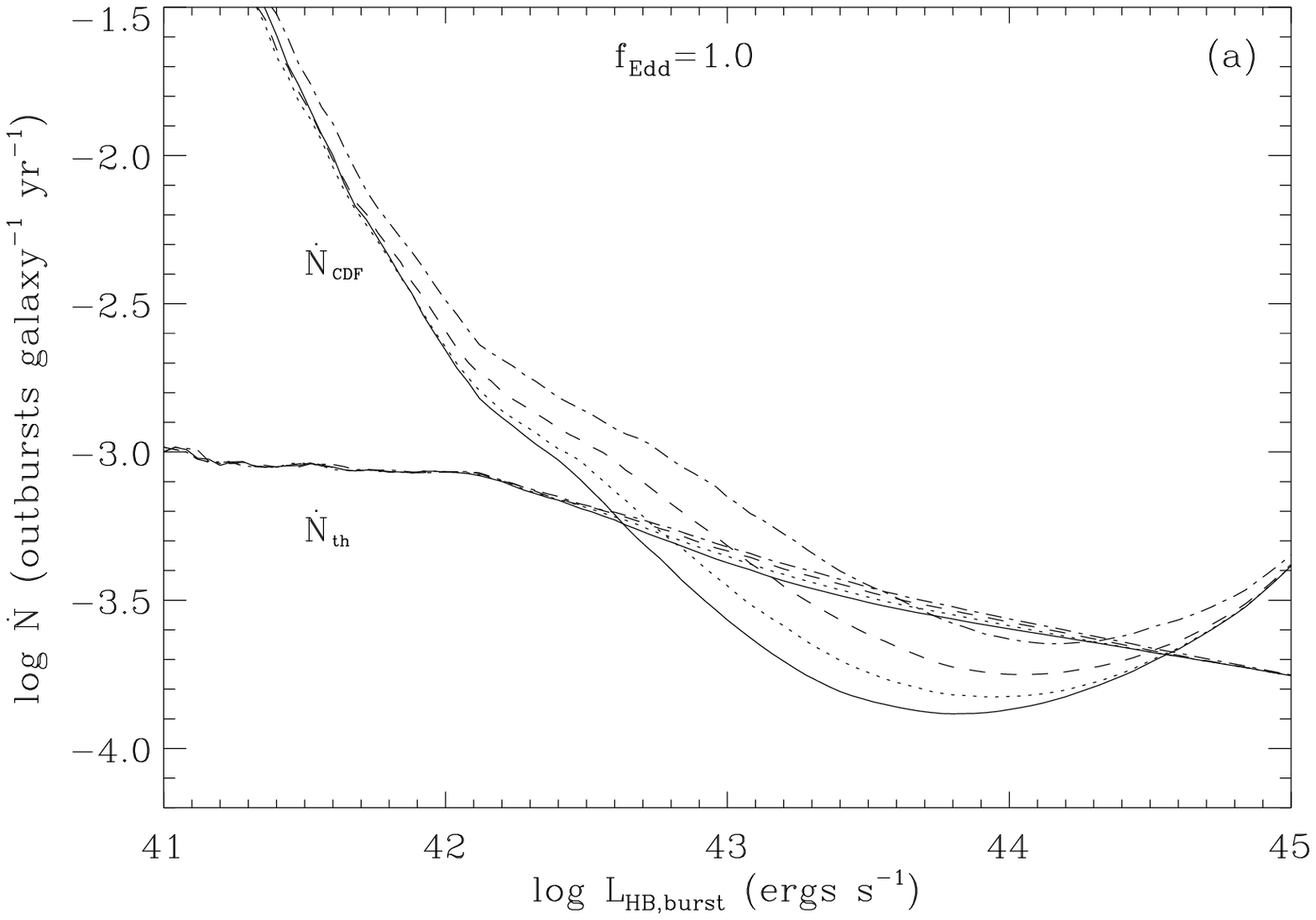}
\includegraphics[scale=0.5]{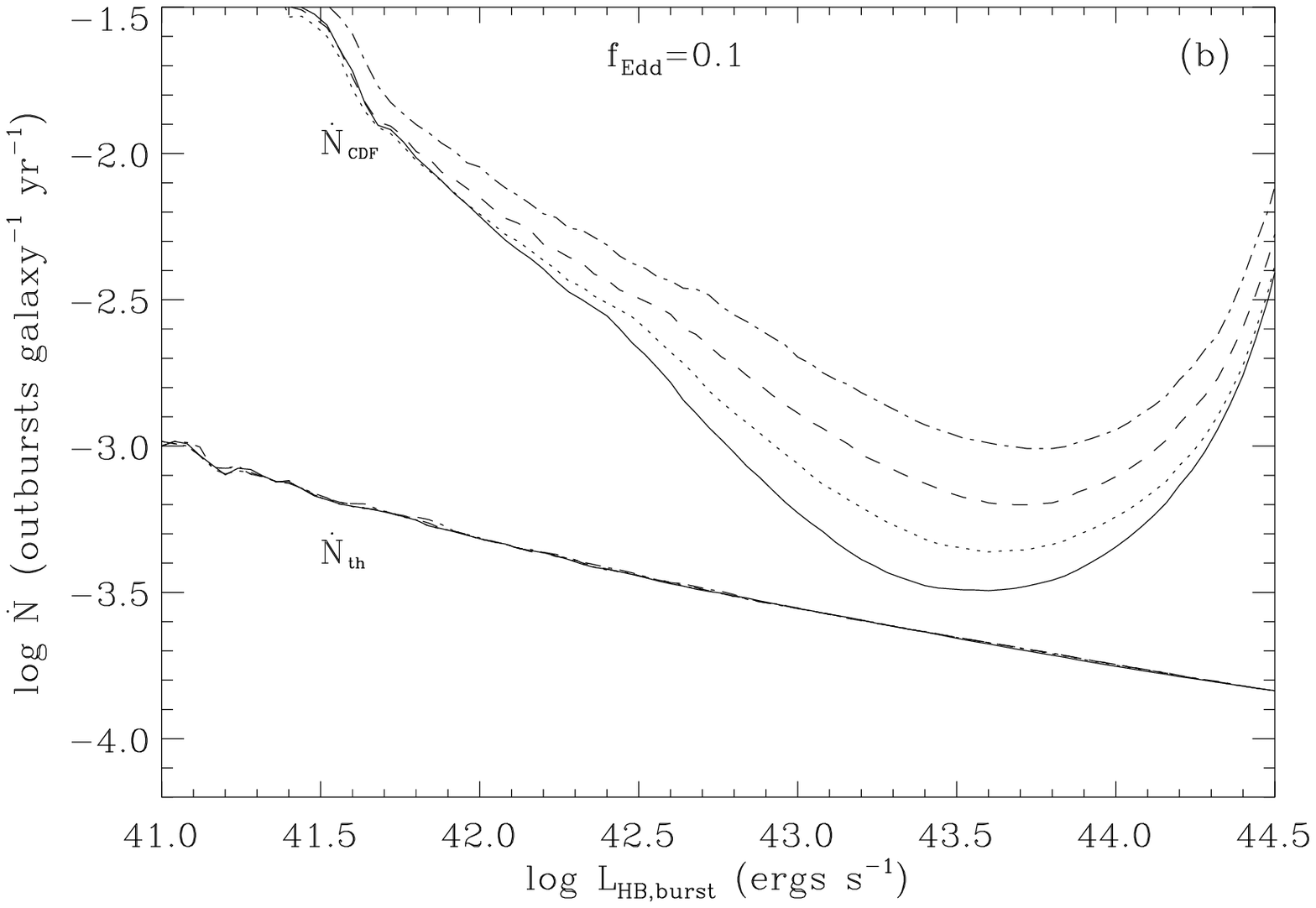}}
\figcaption{The same as Figure \ref{everth}, but for the HB
\hbox{X-ray} luminosity of outbursts. \label{everthhb}}
\end{figure}

As we are here comparing with a physical model, we modified our
previous constraints on event rate by also considering the
capability of a SMBH with a given mass to produce outbursts of
high luminosities. We simply assumed that there was a narrow range
for the outburst luminosity, which was around a fraction $f_{\rm
Edd}$ of the Eddington luminosity. $f_{\rm Edd}=L_{\rm bol}/L_{\rm
Edd}$, where $L_{\rm bol}$ is the outburst's bolometric luminosity
and $L_{\rm Edd}\approx 1.3 \times 10^{38}(M_{\rm
BH}/M_{\sun})$~ergs~${\rm s}^{-1}$ is the Eddington luminosity. We
also assumed a bolometric correction $f_{\rm bc}$, which was
defined as $f_{\rm bc}=L_{\rm bol}/L_{\rm X, burst}$. The
constraint on $M_{\rm BH}$ is then
\begin{equation}
M_{\rm BH}\ge \frac{f_{\rm bc}L_{\rm X, burst}}{1.3
\times 10^{38}f_{\rm Edd}}M_{\sun}.
\label{equ_edd}
\end{equation}
\noindent Applying this SMBH-mass requirement in addition to the
$10^5M_{\sun} < M_{\rm BH} < 3 \times 10^8M_{\sun}$ requirement in
step~2 of \S3.1, we got weaker constraints on the event rate for
high outburst luminosities, since the number of qualified galaxies
$n$ is smaller. The derived upper limits as a function of $L_{\rm
X, burst}$ for the SB and the HB are shown in Figures \ref{everth}
and \ref{everthhb}, respectively. Here we assume $T_{\rm burst}=6$
months, $f_{\rm bc}=10$, and $f_{\rm Edd}=1.0$ or 0.1. The event
rate shows a dependence on the Eddington ratio, $f_{\rm Edd}$,
since smaller $f_{\rm Edd}$ will limit the number of galaxies that
are capable of making bright outbursts. The new requirement on the
SMBH mass does not affect the outburst rate at small $L_{\rm X,
burst}$, because all galaxies are capable of producing outbursts
of such low \hbox{X-ray} luminosity.

To employ the formula for the theoretical tidal-disruption rate of
a single galaxy (Equation \ref{nth}) in our survey, we used
$M_{\rm BH}$ calculated in \S3.1. As the dependence of
$\dot{N}_{\rm WM}$ on $M_{\rm BH}$ is weak, there should not be a
large error if our SMBH-mass estimation is not highly accurate; an
error of a factor of 5 in $M_{\rm BH}$ only changes $\dot{N}_{\rm
WM}$ by a factor of 1.5. As we are only considering tidal
disruptions of stars here, the disruption rate is dominated by
subsolar stars according to the stellar mass function
\citep{Milosavljevic2006}. The radius of these stars follows the
relation $R_{\star}/R_{\sun}\approx (m_{\star}/M_{\sun})^{0.8}$
\citep[e.g.,][]{Kippenhahn1990}. Thus the dependence on stellar
mass and radius is weak, and we simply assumed solar mass and
radius. For each $L_{\rm X, burst}$, we took the average value of
$\dot{N}_{\rm WM}$ for all galaxies for which we are sensitive to
an outburst of luminosity $\ga L_{\rm X, burst}$, which required
$T_{i, {\rm sens}}>0$. The predicted event rate for this survey as
a function of $L_{\rm X, burst}$ is then given by
\begin{equation}
\dot{N}_{\rm th}(L_{\rm X, burst})=\frac{\sum_{i=1}^{n}{k_i\dot{N}_{\rm WM}}}
{\sum_{i=1}^{n}{k_i}} \ \ \
\frac{\rm outbursts}{\rm galaxy\ yr},
\end{equation}
\noindent where $k_i=1$ if $T_{i, {\rm sens}}>0$, and $k_i=0$
if $T_{i, {\rm sens}}=0$.

\begin{figure}
\centerline{\includegraphics[scale=0.5]{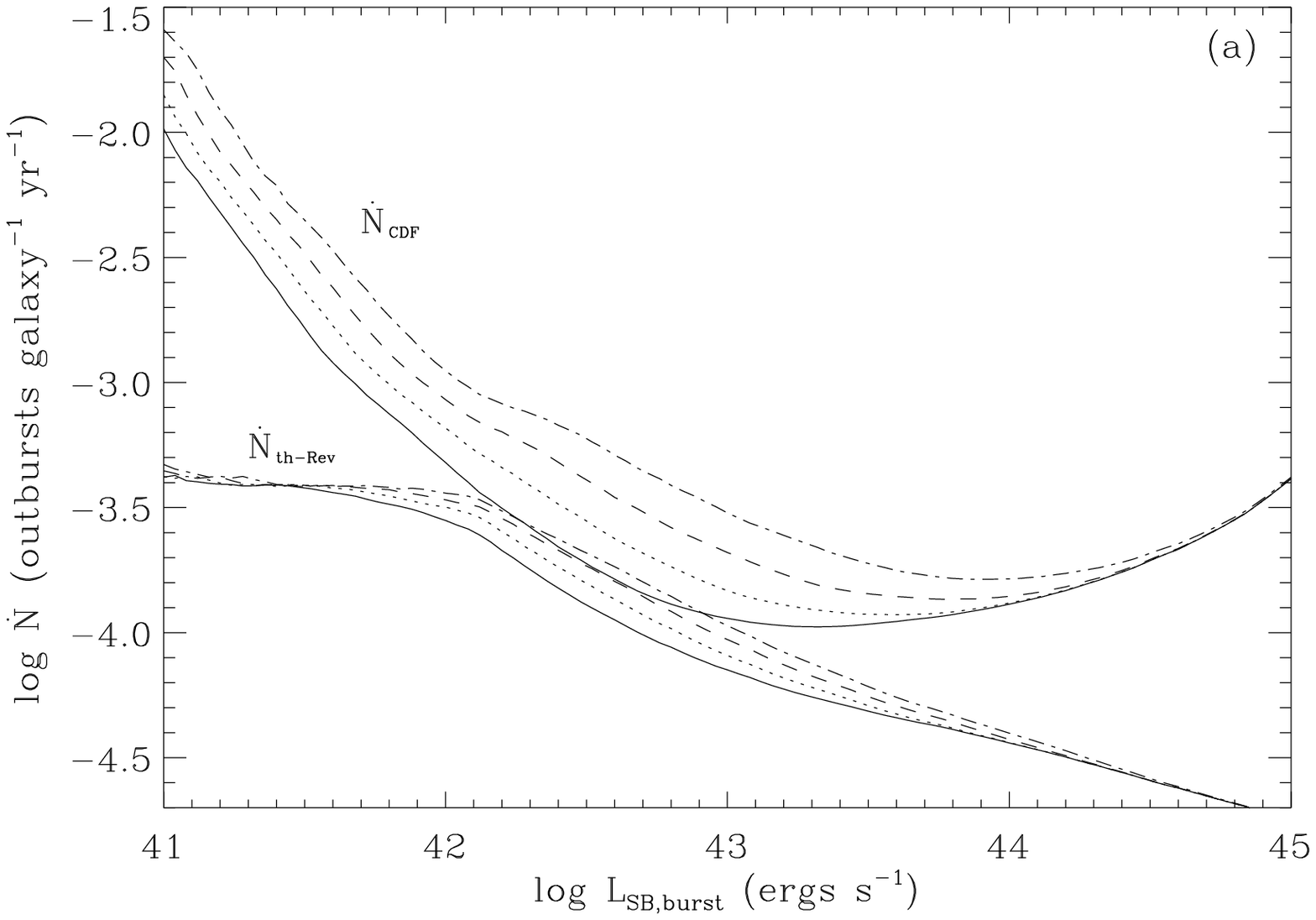}
\includegraphics[scale=0.5]{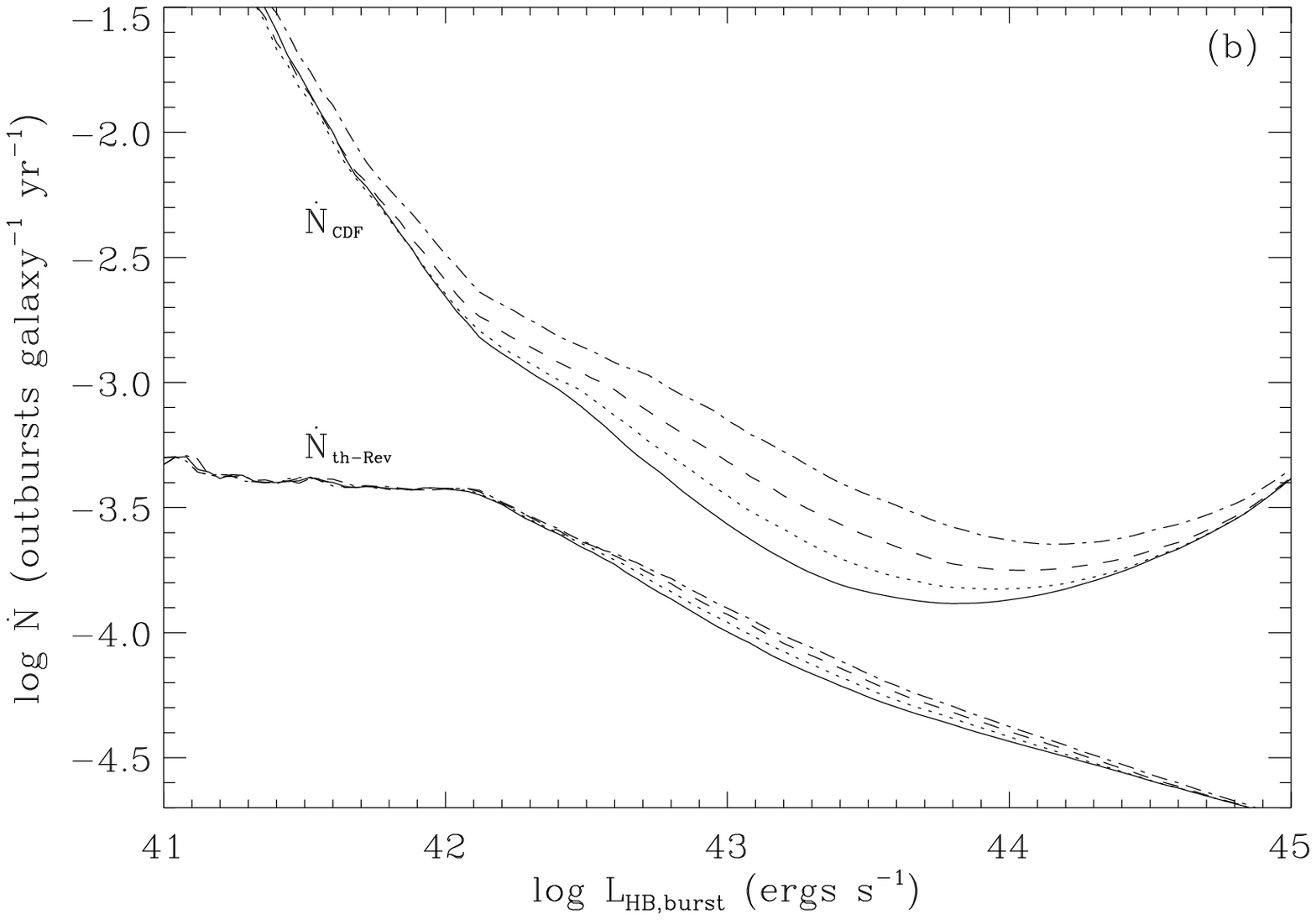}}
\figcaption{90\% confidence upper limits on the event rates of
outbursts derived from this survey and expected event rates from
the theoretical prediction for the ({\it a}) SB and ({\it b}) HB,
under the assumptions of $T_{\rm burst}=6$ months, $f_{\rm bc}=10$
and $f_{\rm Edd}=1.0$. Solid, dotted, dashed, and dash-dotted
lines represent photon indices $\Gamma$ of 2, 3, 4, and 5,
respectively. The observational upper limits are the same as those
in Figures \ref{everth} and \ref{everthhb}, while the predicted
rates were derived from a modified relation which does not require
isothermal stellar density distributions in galactic nuclei
(Equation \ref{nthmod}). \label{everthmod}}
\end{figure}

$\dot{N}_{\rm th}$ as a function of $L_{\rm X, burst}$ for
different spectral shapes and Eddington ratios is also plotted in
Figures~\ref{everth} and \ref{everthhb}. Comparing these
predictions with our survey constraints, we see that the upper
limits derived in this survey are consistent with the predicted
rates, except that around an \hbox{X-ray} luminosity of $10^{43}$
ergs ${\rm s}^{-1}$, this deep-field survey sets a tighter
constraint on the rate of Eddington-limited events than the theory
(Figure \ref{everth}a, Figure \ref{everthhb}a). If we reduce
$f_{\rm Edd}$ to 0.1, the discrepancy decreases and almost
disappears (Figure \ref{everth}b, Figure \ref{everthhb}b). There
are other sources that may help to resolve this discrepancy: (1)
As mentioned in \S1, a tidal-disruption event may not exhibit the
expected outburst characteristics, due to a short duration of the
event or \hbox{X-ray} obscuration. It is then likely to be missed
by current and previous surveys. (2) Equation \ref{nth0} is
derived under the assumption that the galactic nucleus is a
singular isothermal sphere. For other kinds of density
distributions, the predicted rate will become smaller according to
Figure 5b of \citet{Wang2004}, which gives the computed
tidal-disruption rates and SMBH masses for all the galaxies in
their sample. A straight-line fit to the data points in this plot
gives
\begin{equation}
\dot{N}_{\rm WM-Rev}(M_{\rm BH})\approx 2.3 \times 10^{-4}\ {\rm yr}^{-1}
\left(\frac{M_{\rm BH}}
{10^6M_{\sun}} \right)^{-0.52},
\label{nthmod}
\end{equation}
with a smaller rate and stronger dependence on SMBH mass. A
comparison with the predicted tidal-disruption rate derived from
this modified relation is shown in Figure \ref{everthmod}. The
revised rates $\dot{N}_{\rm th-Rev}$ are $\sim0.5$ dex smaller and
now are slightly below our observed upper limits for the full
luminosity range. (3) The $M_{\rm BH}$-$\sigma$ relation in
Equation \ref{msig} only holds for local SMBHs. The relation could
evolve in the sense of velocity dispersion decreasing
\citep{Woo2006} with redshift for a fixed BH mass, making the
predicted rate decrease at high redshift. All these effects may
cause the upper limits on event rate given by this survey to be
above the predicted stellar tidal-disruption rates. On the other
hand, the galaxy sample in \citet{Wang2004} is a set of 61
elliptical galaxies, while our sample is dominated by late-type
galaxies. By considering the contribution from the bulges of
spirals, the predicted rates could be significantly increased
\citep{Wang2004}, as the masses of the bulges are smaller than the
total masses of the galaxies that we were using. This could
increase the discrepancy between the observations and theoretical
predictions.

\begin{figure}
\centerline{\includegraphics[scale=0.5]{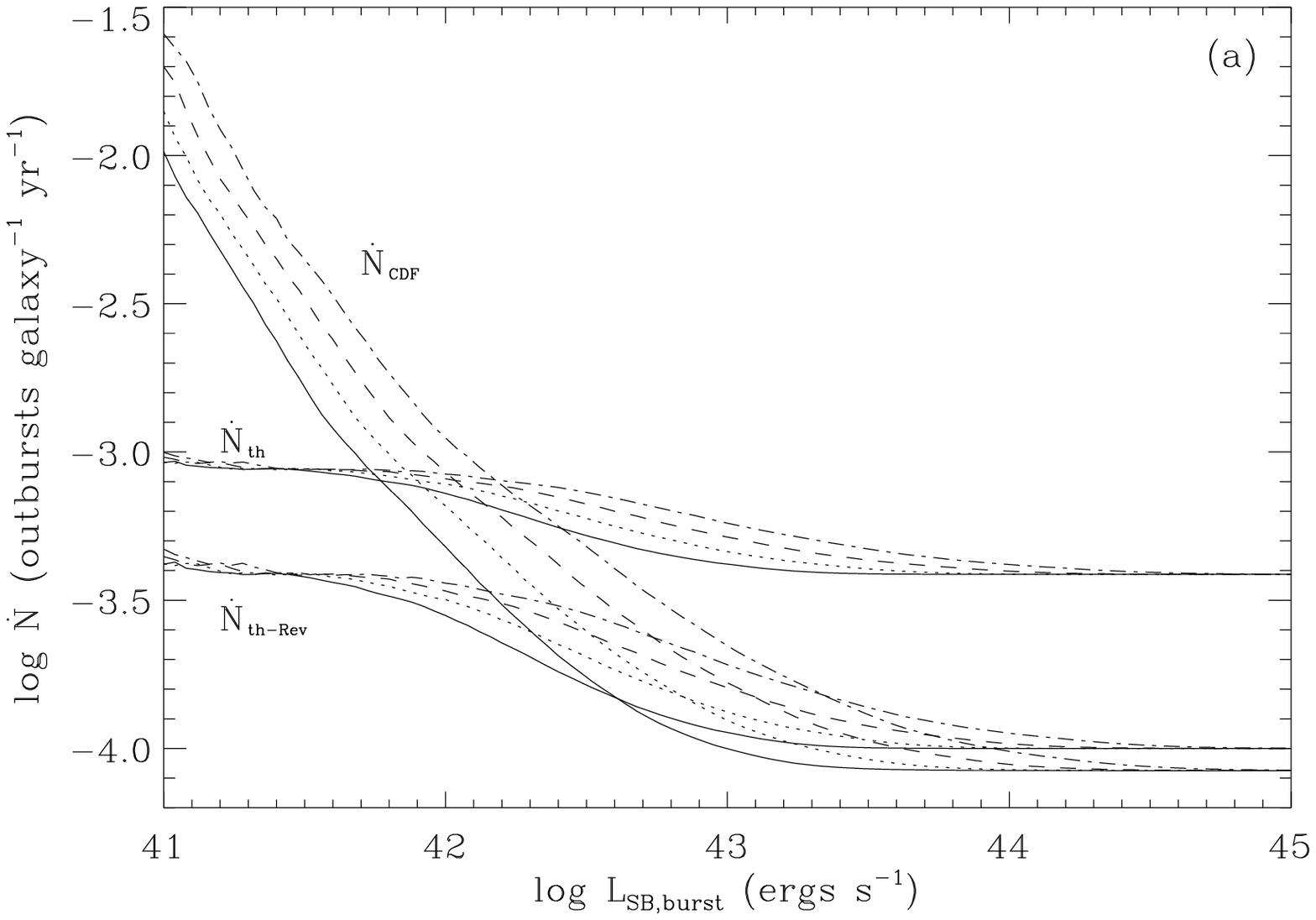}
\includegraphics[scale=0.5]{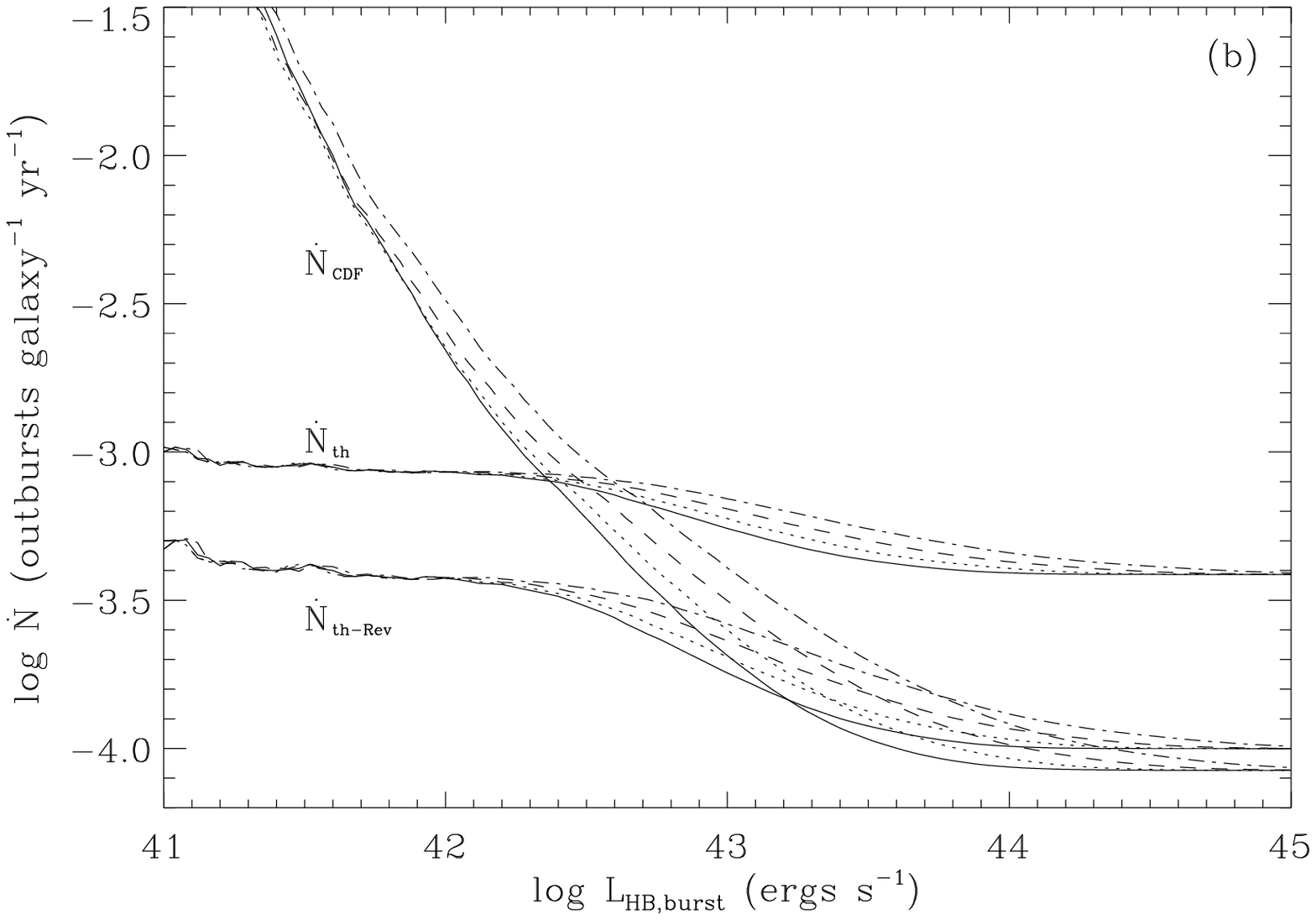}}
\figcaption{90\% confidence upper limits on the event rates of
outbursts derived from this survey and expected event rates from
the theoretical prediction for the ({\it a}) SB and ({\it b}) HB,
under the assumption of $T_{\rm burst}=6$ months. Solid, dotted,
dashed, and dash-dotted lines represent photon indices $\Gamma$ of
2, 3, 4, and 5, respectively. The \hbox{X-ray} luminosities were
not constrained by the Eddington limit in these plots. The
observational upper limits are the same as those in Figure
\ref{ever}. The predicted rates were derived from Equation
\ref{nth} which is the analytic relation for galactic nuclei with
isothermal stellar density distributions, and from Equation
\ref{nthmod} which does not require isothermal stellar density
distributions. \label{everthnoedd}}
\end{figure}

We employed the Eddington limit above based on the assumption that
the \hbox{X-rays} come from a transient accretion disk and its
corona. Alternatively, \hbox{X-rays} could instead come from
tidal-stream collisions \citep[e.g.,][]{Kochanek1994}, in which
case the luminosity could perhaps exceed the Eddington limit. If
we remove the constraint on the SMBH mass set by
Equation~\ref{equ_edd}, the upper limits on the event rates get
tighter for high \hbox{X-ray} luminosities and are the same as
those in Figure~\ref{ever}, and the predicted rates also change
due to their dependence on $M_{\rm BH}$. The results are plotted
in Figure \ref{everthnoedd}, which shows the predicted rates from
both the original relation (Equation \ref{nth}) and its revision
(Equation \ref{nthmod}). Our survey constraints for $L_{\rm X,
burst}\ga 10^{43}$~ergs~s$^{-1}$ are $\sim 0.5$ dex tighter than
the analytical predictions assuming isothermal stellar density
distributions in galactic nuclei, and are consistent with the
revised tidal-disruption rates derived from the full galaxy sample
in \citet{Wang2004}.

We have assumed $A(z)=1$ in the above analyses, which represents
no redshift evolution of the outburst rate. However, the rate of
\hbox{X-ray} outbursts could increase with redshift owing to
changes in galactic nuclei and SMBH masses.
\citet{Milosavljevic2006} have proposed that stellar tidal
disruptions are largely responsible for the AGN \hbox{X-ray}
luminosity function at luminosities below
\hbox{$10^{43}-10^{44}$}~erg~s$^{-1}$. If this is indeed the case,
then the order-of-magnitude increase in the comoving number
density of such AGNs out to $z\sim 1$ (e.g., Brandt \& Hasinger
2005 and references therein) implies that the outburst rate must
correspondingly increase. Scaling the Wang \& Merritt (2004) rates
upward by $A(z=0.8)\approx 10$ to account for the median redshift
of our sample would lead to significant disagreement with our
observational constraints, even if the revised rate (Equation
\ref{nthmod}) is adopted, suggesting that the \hbox{X-ray}
luminosity function at luminosities below
$10^{43}-10^{44}$~erg~s$^{-1}$ is not primarily due to stellar
tidal disruptions.

On the observational side, Donley et~al. (2002) performed a
systematic survey for \hbox{X-ray} outbursts using the {\it ROSAT}
database (although this survey had some substantial systematic
uncertainties owing to complex selection effects). They detected
five outbursts and placed the first constraints on the rate of
such outbursts. The rate of large-amplitude \hbox{X-ray} outbursts
from inactive galaxies in the local Universe is \hbox{$\sim
10^{-5}$~galaxy$^{-1}$~yr$^{-1}$}, estimated from the survey
volume and the galaxy space density. Compared to their results,
our survey constraints are based on uniform observational data for
$24\,668$ optical galaxies which are less biased, and there are no
uncertainties introduced in the estimation of survey volume or the
galaxy space density. Moreover, we are able to probe \hbox{X-ray}
outbursts with higher redshifts and in a harder \hbox{X-ray} band
for the first time, and we derived luminosity-dependent rate
constraints which offer insight into the low-luminosity regime. HB
\hbox{X-ray} surveys are able to detect obscured outbursts which
could have been missed in previous surveys; intrinsic column
densities as low as $\sim 5\times 10^{21}$~cm$^{-2}$, perhaps
associated with gas from the tidally disrupted star, would greatly
reduce the detectability of soft-spectrum outbursts in the \rosat\
band. Here we adopted $T_{\rm burst}=6$ months, the same as in
Donley et~al. (2002). The SB and HB constraints are shown in
Figure \ref{ever}. From $L_{\rm X, burst}\sim 10^{43}$ down to
$10^{41}$ ergs~s$^{-1}$, the upper limit on the event rate
increases from $\sim 10^{-4}$~galaxy$^{-1}$~yr$^{-1}$ to $\sim
10^{-2}$~galaxy$^{-1}$~yr$^{-1}$, mainly due to the limited
sensitivity in the low-luminosity regime. Further observations are
required to assess whether low-amplitude events, which could come
from partial tidal disruptions or the accretion of brown dwarfs,
planets, or small gas clouds, are truly more common than
high-amplitude events, as for other transient phenomena in nature.
When $L_{\rm X, burst}\ga 10^{43}$~ergs~${\rm s}^{-1}$, the upper
limit on the event rate is $\sim 10^{-4}$~galaxy$^{-1}$~yr$^{-1}$,
either in the SB or the HB. Compared to the results in Donley
et~al. (2002), these constraints allow an amplification of the
outburst rate up to a factor of 10 at most when considering
obscured \hbox{X-ray} outbursts and redshift evolution from $z\sim
0$ to $z\sim 0.8$.

\section{CONCLUSIONS AND FUTURE WORK}

In summary, we constructed a sample of $24\,668$ optical galaxies
in the \hbox{CDF-N} and \hbox{CDF-S} with redshifts obtained from
the literature; the median redshift is $\sim 0.8$. We analyzed
exceptionally sensitive \chandra\ observations of these galaxies,
which span $798$ days for the \hbox{CDF-N} and $1\,828$ days for
the \hbox{CDF-S}. We searched for \hbox{X-ray} outbursts with the
criterion that the count rate varies by a minimum factor of 20 in
one of three standard bands. No outbursts were found, and thus we
set upper limits on the rate of such events in the Universe, which
depend on the \hbox{X-ray} luminosity of outbursts. If we only
consider those galaxies hosting SMBHs, and those with SMBHs not
massive enough to swallow a whole star without disruption, we
derive an upper limit on the rate of an outburst with $L_{\rm X,
burst}\ga 10^{43}$~ergs~${\rm s}^{-1}$ and $T_{\rm burst}=6$
months to be $\sim 10^{-4}$~galaxy$^{-1}$~yr$^{-1}$ (Figure
\ref{ever}), without any other assumptions about the physical
model producing \hbox{X-ray} outbursts. Compared to the survey by
Donley et~al. (2002), our survey probes both higher redshifts and
harder \hbox{X-ray} energies. The outburst rate may increase by a
maximum factor of 10 when taking into account both obscured
\hbox{X-ray} outbursts and redshift evolution from $z\sim 0$ to
$z\sim 0.8$. We are able to set constraints on low-luminosity
events down to $10^{41}$~ergs~s$^{-1}$, though at this luminosity
the event rate is limited by sensitivity.

We also compared our constraints to the predicted tidal-disruption
rates of \citet{Wang2004}, exploring several possibilities in
physical parameter space. If the outburst luminosity is limited by
the Eddington luminosity with $f_{\rm Edd}\la 0.1$, or the
predicted rate is computed from a modified relation which does not
require isothermal stellar density distributions in galactic
nuclei, our results are roughly consistent with theoretical
predictions. Otherwise, our constraints are tighter than the
predictions. Moreover, if stellar tidal disruptions are largely
responsible for the AGN \hbox{X-ray} luminosity function at
luminosities below \hbox{$10^{43}-10^{44}$}~erg~s$^{-1}$ as
proposed by \citet{Milosavljevic2006}, the predicted rate should
be scaled upward by $A(z=0.8)\approx 10$ to account for the median
redshift of our sample. This scaling leads to significant
discrepancies with our observational constraints, suggesting that
the \hbox{X-ray} luminosity function at low luminosities is not
likely to be dominated by stellar tidal disruptions.

The constraints set by this study could be significantly improved
by further deep-field surveys and new missions. For example, the 1
Ms of additional \hbox{CDF-S} exposure starting in 2007 September
will increase the monitoring baseline significantly and provide
$\sim 13$ new observations which could be grouped into several
additional epochs for outburst searching and other variability
studies. Future missions such as {\it Lobster\/} and {\it
eROSITA\/} in the soft \hbox{X-ray} band and the {\it BHFP\/} in
the hard \hbox{X-ray} band will also be capable of detecting and
studying outbursts, benefiting from their large sky coverage.
\citet{Grindlay2004} predicted that the {\it Energetic
\hbox{X-ray} Imaging Survey Telescope (EXIST)} implementation of
the {\it BHFP\/} would detect \hbox{X-ray} outbursts out to $\sim
100$~Mpc at a rate of $\sim 30$~yr$^{-1}$, adopting the rates of
\citet{Wang2004}. Such missions will put better constraints on the
outburst rate, the fraction of obscured outbursts, and redshift
evolution, enhancing our knowledge about the nature of
\hbox{X-ray} outbursts.

\section{ACKNOWLEDGEMENTS}
We thank P. Capak, B. D. Lehmer, A. Marconi and D. Merritt
for helpful discussions. We acknowledge financial support from
\chandra\ \hbox{X-ray} Center grants AR6-7021X (BL, WNB, ATS) and
G04-5157A (BL, WNB, ATS).

\clearpage


\end{document}